\documentclass[screen]{acmart}

\AtBeginDocument{%
  \providecommand\BibTeX{{%
    \normalfont B\kern-0.5em{\scshape i\kern-0.25em b}\kern-0.8em\TeX}}}

\setcopyright{acmcopyright}
\copyrightyear{2025}
\acmYear{2025}

\acmConference[CHI '25]{CHI conference on Human Factors in Computing Systems}{April 26 -- May 01,
  2025}{Yokohama, Japan}
\acmISBN{978-1-4503-XXXX-X/18/06}
\acmBooktitle{CHI 2025} 
\acmISBN{2025}

\usepackage{colortbl}
\usepackage{algorithm}
\usepackage{algpseudocode}
\usepackage{soul}
\makeatletter
\newcount\SOUL@minus 
\makeatother
\usepackage{graphicx}
\usepackage[export]{adjustbox} 

\begin{document}

\title{Actionable AI: Enabling Non Experts to Understand and Configure AI Systems}



\author{Cecile Boulard}
\email{cecile.boulard@naverlabs.com}
\orcid{0000-0002-2188-287X}
\affiliation{%
  \institution{Naver Labs Europe}
  \city{Meylan}
  \country{France}
}

\author{Sruthi Viswanathan}
\email{sruthi.viswanathan@cs.ox.ac.uk}
\orcid{0000-0002-1113-7171}
\affiliation{%
  \institution{University of Oxford}
  \city{Oxford}
  \country{United Kingdom}
  }

\author{Wanda Fey}
\email{wanda.fey-intern@naverlabs.com}
\affiliation{%
  \institution{Naver Labs Europe}
  \city{Meylan}
  \country{France}
}

\author{Thierry Jacquin}
\email{Thierry.jacquin@altai-ego.com}
\affiliation{%
  \institution{AltAI ego}
  \city{Grenoble}
  \country{France}
}

\renewcommand{\shortauthors}{Boulard et al.}

\begin{abstract}

 Interaction between humans and AI systems raises the question of how people understand AI systems. This has been addressed with explainable AI, the interpretability arising from users' domain expertise, or collaborating with AI in a stable environment. In the absence of these elements, we discuss designing \textit{Actionable AI}, 
which allows non-experts to configure black-box agents. In this paper, we experiment with an AI-powered cartpole game and observe 22 pairs of participants to configure it via direct manipulation. Our findings suggest that, in uncertain conditions, non-experts were able to achieve good levels of performance. By influencing the behaviour of the agent, they exhibited an operational understanding of it, which proved sufficient to reach their goals. Based on this, we derive implications for designing Actionable AI systems. In conclusion, we propose Actionable AI as a way to open access to AI-based agents, giving end users the agency to influence such agents towards their own goals.
\end{abstract}



\begin{teaserfigure}
  \includegraphics[width=0.85\textwidth]{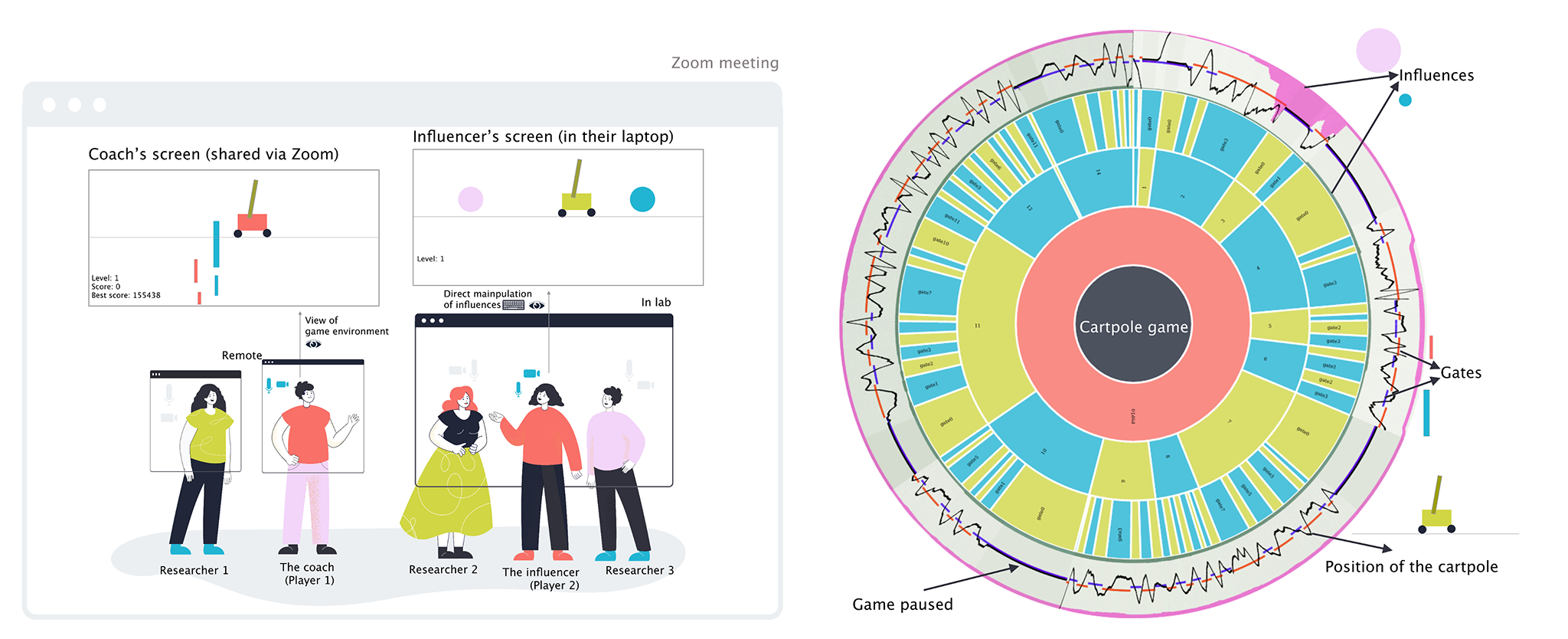}
  \caption{Hybrid experiment setup for non-expert participants to experience \textit{Actionable AI} via a cartpole game (Left) and visualisation of the experiment session of participants P9 and P10 (Right).}
  \Description{Experiment setup and game vis.}
  \label{fig:teaser}
\end{teaserfigure}

\maketitle

\section{Introduction}
`What goes up, must come down.' While physicists might describe gravity as one of the most misunderstood forces of nature \cite{gonen2008study}, a layperson can articulate a sufficient understanding with this common phrase. Similarly, while we may not have an intricate understanding of everyday concepts such as gravity, internet, and banking (as experts in these respective fields do), as everyday users we are able to interact with these systems successfully to get the outcomes we desire. In a similar light, users of AI systems can often come to be everyday users who are neither Machine Learning (ML) experts, nor hold expertise in any particular domain (such as medicine or finance). One of the key aspects of Human-Centred AI (HCAI) research is ensuring control of AI systems to all kinds of users \cite{shneiderman2022human} despite their different levels of knowledge and understanding. Designing user-friendly AI systems that are easy to adapt and adopt is the end game for the success of AI in our society \cite{dourish2003adoptadapt}. 

Today, HCAI efforts aid ML experts, who build, train, and test their ML models, with paradigms such as human-in-the-loop \cite{Wu_2022}, to learn from the domain experts or everyday users. Domain experts without advanced ML knowledge, are supported with interpretable \cite{fiebrink2009meta} and explainable \cite{gunning2019xai} elements to collaborate with AI systems as per their specific needs. Everyday users who have little to no experience with ML, are mostly offered explanations \cite{https://doi.org/10.48550/arxiv.1710.00794} and more recently virtual exploration playgrounds \cite{dang2022prompt} for interacting with AI systems. 

However, the success of these concepts largely relies on either or all of the following: (i) the opportunity to instruct users with explanations \cite{arrieta2020explainable}, (ii) users' prior knowledge of ML \cite{IML} or domain \cite{IMT}, (iii) collaborations with AI in controlled environments \cite{transferability}, and (iv) AI systems built on transparent models \cite{chao2010transparent}. Everyday users do not engage with Explainable AI \cite{miller2023explainableaideadlong}.  In real-life scenarios, embedded AI in autonomous \textit{navigating} systems, such as autonomous robots questions these expertise-, explanation-, or environment-dependent HCAI concepts. 
Having an autonomous robot in houses or large buildings require the ability to adapt its behaviour to ensure performance and safety. In such personal environments, risks, expectations, norms, culture, habits, language are so specific that success of domestic autonomous robots is bounded to an easy way to configure it \cite{Ackerman}. Rethinking and pre-planning affordances for embedded AI scenarios, at different levels of system, environment, and observer (user) \cite{Planningaffordance} is a need of the hour.


We propose the framework of \textit{Actionable AI}, for proactively instilling affordances in AI systems for interactions with everyday (non-expert) users, facing unexplained (blackbox) models, in uncertain environments. In this contribution, we focused on experimenting the feasibility  of this Actionable AI framework, answering the following research question.



\textit{Research Question: Can non-experts succeed in configuring a black-box agent in an actionable environment to meet their goals without a prior understanding of the rules and goals of the agent?}

To answer our research question, we bring about Actionable AI, by building on: (i) the environment of real-time strategy video games \cite{RTS}, (ii) a Human-in-the-loop technique called \textit{influences} \cite{JacquinPerezBoulard}, and (iii) Open AI Gym's cartpole agent \cite{Cartpole}. In the classic cartpole game \cite{OpenAIGym}, a pole (\textit{reversed pendulum}) placed on the top of a horizontally moving cart, is controlled by an AI algorithm that tries to keep the pole from falling off as long as possible. In our adapted version of this game (Figure \ref{fig:teaser}), players had to move the cartpole in the screen to pass all the gates (vertical lines as in a pseudo ski slalom) using influences (two circles on the screen controlled by the keyboard). Our experimental setup was a black box for non-expert participants who received no explanations, neither about the cartpole, nor about the influences. We ran the experiment, with 30-minute gaming time, in a hybrid setting with 22 teams of two players each (44 participants). Half of the teams (22 out of 44 participants), were informed about the AI's control of the cartpole before the experiment, while the rest were informed after. For every session, one participant was physically co-located with us and the other was remote. This organisation allowed us to mimic the context of collaborative games.


Our findings suggest that 14 out of 22 teams were able to reach a good level of performance in the AI-controlled cartpole and its influences, as they could configure its behaviour for winning the cartpole game with unknown rules. Indeed, their learning curves were fast, and the use of influences was intuitive. 20 out of 22 players who used the influences to control the cartpole developed an understanding of the system that led them to acquire operating knowledge, which proved sufficient for interacting with the cartpole black box. Of the 22 participants who were informed about the presence of an AI in the system, 15 self-reported that they failed to remember it during their interaction. Despite this, they were still able to act on the AI and configure it to meet their needs, without necessarily needing to fully understand its underlying autonomy. This suggests that the participants focused on their ability to adapt/control the AI, rather than the fact that it was autonomous. Our observations also unveil how influences gave way to the \textit{affordance} of different strategies created by the participants during the course of the game, such as securing the cartpole's environment, efficiently moving the cartpole, or reacting swiftly to manage the situation when the model fails. These results suggest that the implementation of Actionable AI in our contribution, enabling non-experts to use influences to configure ML models, efficiently supports the real-time tuning of AI systems. As implications for design, we elaborate on the key elements of the Actionable AI framework. Stemming from these findings we discuss the evidence of successful Human-AI collaboration and the different levels of explanations unravelled in our study.


In summary, to enable non-experts to configure AI systems, we introduce \textit{Actionable AI}, a framework for Human-AI interaction, which is grounded on the use of technological means that allow direct interaction with an AI agent. We build a reinforcement-learning based cartpole game (Section \ref{sec:prototype}), by leveraging the technique of \textit{influences}. With our modified cartpole game, we experiment the feasibility of Actionable AI with non-experts without providing them any instructions to deal with our black-boxed AI (Section \ref{sec:method}). Our findings suggest that, via exploring the direct manipulation of an unknown AI agent, participants gain operational knowledge of it, and succeed to configure it towards achieving their goals (Section \ref{sec:findings}). We develop and discuss the potential of Actionable AI to support human-AI interaction aligned with HCAI concerns (Sections \ref{sec:ImplDesign} and \ref{sec:DiscExplanation}). The main contributions of this work include:

\begin{itemize}

\item Empirical findings for designing \textit{Actionable AI} for non-expert users to empower them to configure black-box AI systems in uncertain situations.
\item Through our real-time gaming experiment, containing an AI agent connected to interactive \textit{influences}, we verify non-experts' ability to succeed in discovering the system's affordances and to come up with game-winning strategies in a limited amount of time.
\item Finally, we derive implications for designing \textit{Actionable-AI} systems, with direct manipulation, a visible action space, some time to experiment, levels of learning, and performance indicators.
\end{itemize}



\section{Related Work}

\subsection{Users understanding AI}

Explainable AI is one of the most explored methods in the HCI literature for several application categories of Human-AI interaction. AI stakeholders' key motivations for building explainable systems include debugging, identifying bias, and building trust \cite{10.1145/3334480.3383047}. Explainability has also been used to promote data transparency thus allowing the end users to assess the ML system's fairness \cite{10.1145/3334480.3383047}. While explainable AI for domain experts (such as doctors) who can interpret the reasoning of ML models can be achieved in a straightforward manner using the system's variables (Interpretable AI), non-experts (such as patients) need user-driven designing and additional variables to help them understand AI (comprehensible AI) \cite{https://doi.org/10.48550/arxiv.1710.00794}. Non-experts do not have the agency to engage with explanations for AI's decision making \cite{miller2023explainableaideadlong}. As automation increases, it is also important to increase human understanding and control of AI systems to ensure the building of human-centred and responsible AI \cite{shneiderman2022human}.

In particular, AI black boxes such as systems built with deep neural networks cannot afford direct manipulation by non-expert end users, who have no or limited knowledge of Machine Learning. While explainability is still useful for operating task-based AI systems such as domestic Robots \cite{RobotExplanation}. This gets increasingly difficult and complicated when future black box AI systems are tasked to deliver explanations in real-world scenarios \cite{8490530}, where systems will be deployed in public environments around multiple users. AI is even deemed unexplainable and incomprehensible in various real-world situations, also adding a critical trade-off between the system's ability to provide either an optimal decision or an explainable/comprehensible one\cite{unexplainable}. Yet, it is essential to maintain end-user control over the AI system for AI to be successful, useful, and usable when deployed in the real world \cite{shneiderman2022human}. Therefore, it is evident that human-AI interaction has to venture beyond explainable AI to facilitate user control over unexplainable AI systems of the future.

\subsection{Users improving AI}
Human-in-the-loop (HITL) is a notable paradigm employed by ML researchers and practitioners for optimising algorithms with the help of end users\cite{Wu_2022, https://doi.org/10.48550/arxiv.1909.09906}. In various AI domains such as computer vision and natural language processing, by using HITL, non-experts help improve AI with data processing, data annotation, and iterative labeling of data. As described in \cite{Wu_2022}, human guidance can be used during the learning phase of the agent with various approaches. The agent can learn by imitation (of the action, or the outcome of the action, i.e the environment), with human feedback (evaluation, preference, or attention). The agent can also receive hierarchical goals from the human teacher. HITL literature groups all kinds of \textit{human guidance}, not addressing the role of the human. Is the feedback or the teaching active (to give a preference or an evaluation) or passive (agent relying on the human gaze or outcome of its action)? More active methods for human feedback include techniques such as Active learning, which is used to interactively involve domain experts, crowd workers, or end users by designing queries for them to label a selectively leveraged set of data \cite{settles2009active}. Using Interactive Machine Learning (IML), given the interaction opportunity usually by involving a user interface, non-experts refine ML models by adding their input and reviewing the changes in an iterative fashion \cite{IML}. For IML to be efficient, users' mental model of how the system works must be well aligned with the deployed ML algorithm \cite{IMLDebugging}. Interactive Machine Teaching (IMT) envisions leveraging the inherent teaching  capabilities of humans to facilitate non-experts to teach to ML models\cite{IMT}. This also involves providing specific guidance to non-experts and other challenges such as designing relevant user experiences and teaching languages. All the above-discussed longitudinal techniques allow non-experts to modify the model over an extended time by annotating samples and reviewing their impact on the model. While it is essential to involve users in iterative training of models, which primarily serves only to improve the training of the model. We also identify the need to enable end users to configure model outcomes and thereby tune the behaviour of the system for immediate use in real-world scenarios. 

Real-time adaptation and usage of AI systems, which are closer to our goal, have been achieved by a lesser number of works in the literature. In constructing a meta-instrument called Wekinator \cite{fiebrink2009meta}, musicians successfully trained and used an ML model to exhibit on-the-fly learning. With a  transparent model, the active robot learning \cite{cakmak2010designing} method allows everyday users to actively train robots in real-time. Situational recommender \cite{SitRec} allows users to control recommendations using sliders in the UI from their first use. However, these paradigms are limited to domain experts \cite{fiebrink2009meta}, or rely on the transparency of the ML models \cite{cakmak2010designing}, or need an extensive domain- and context-based user research to classify different situations of use \cite{SitRec}.
 



\subsection{Users playing with AI}
Deploying AI in the real world involves creating robust models which adapt to unforeseen contexts and ultimately designing for successful Human-AI collaborations, which holds the key to increasing the willingness of end users to adopt AI into their work and life \cite{dourish2003adoptadapt}. The literature on Human-AI collaboration focuses on various aspects of increased task performance of Human-AI teams \cite{Gao2021HumanAICW},  user perception of AI \cite{userperceptions}, different levels of authority in the roles assigned to AI \cite{collabRoles}, and biases sustained by users when collaborating with AI vs. human \cite{directionality}. Another important aspect of human-AI collaboration is communication and coordination \cite{liang2019implicit}. In many cases, humans and AI systems need to be able to share information and coordinate their actions to work together effectively \cite{wang2020human}. This can involve the use of specialized tools and interfaces that facilitate communication and coordination between the two \cite{directionality,wang2020human,Gao2021HumanAICW}.

Recently the case of transferability \cite{transferability} of AI, from researchers to end users, has been focused as an enabler of human-ai teaming. Where researchers propose to test communication and collaboration in an environment such as a Real-time strategy (RTS) video game to learn lessons that can be applied to real-world situations \cite{transferability}. As RTS games often include elements that are relevant to real-world scenarios, such as resource management, decision-making under uncertainty, and collaboration \cite{RTS};  they have been ideal to test situational interaction, decision-making, and awareness in different contexts \cite{transferability}.  Also, video games are associated with promoting fluid intelligence \cite{fluidIntel},  which is the ability to think and reason abstractly and solve novel problems, independent of any specific knowledge or experience. 

 While the above example \cite{transferability}  promotes end users testing AI via RTS games to unearth possible pitfalls in a moderated environment, recent success of virtual playgrounds for large language models such as DALL·E 2 \cite{Dalle2} and ChatGPT \cite{chatgpt}, has pushed the use of game-like environments from testing AI to exploring and using it. ChatGPT attracted over a million users in several days \cite{chatgptUsers}, however, people playing with are to a large degree uncertain about the outcome of their inputs, even after exploring and re-exploring the model many times \cite{atlantic}.

\section{Methodology}
\label{sec:method}



\subsection{The cartpole game}

 We designed a cartpole game similar to Slalom skiing, an alpine skiing sport, where skiers make quick turns to avoid vertically suspended blue- and red-coloured poles/gates. Similarly, in our game players have to place the cartpole to the right of blue gates and then place it to the left of red gates. Players win the game when all gates are successfully crossed by alternating the cartpole's position to the left and right. Players lose the game either when the cartpole exit the screen or when the pole falls off the cart.
 We designed three levels in the game which gradually add complexity to the game (see Table \ref{tab:level}). The added disturbances in the environment, namely slope, wind, and bumps, have a mechanical effect on the cartpole. The slope of the road (the line on which the cartpole moves), creates a gravity-like effect on the cartpole that causes it rolls down in the slope. The wind disturbance creates wind gusts that pushes the pole off the cart. The bumps disturbance creates a bumpy road by slightly varying the height of the road, similar to creating speed bumps in the path of the cartpole.

\begin{table}
    \centering
        \begin{tabular}{l|ccc} 
            \hline
            \textbf{Level} & \textbf{Gates} & \textbf{Disturbances} & \textbf{Type of Disturbances} \\
            \hline
            \textbf{1} & 4 & 0 & none \\
            \textbf{2} & 8 & 1 & slope and wind \\
            \textbf{3} & 12 & 2 & slope, wind, and bumps \\
             \hline
        \end{tabular}
    \caption{The description of the game levels with the number of gates that needs to be consecutively achieved to win the level, the number of environmental disturbances on gates and the type of disturbances}
    \label{tab:level}

\end{table}

To pass a level and progress to the next, players have to win three consecutive games in that level. This three consecutive wins rule was designed to avoid the effect of players getting lucky in the game and allow participants to truly experiment with the effect of the influences on the cartpole before facing more-difficult situations with added disturbances in the environment at the next levels.

\subsection{Experimental design}

To foster natural verbalisation from the players, we transformed the one-player cartpole game into a two-player cartpole game with partial information. Meaning that each player has access to a part of the information to win the game. Both players had to collaborate and talk together to score in the game. Our target was to unearth a collection of topics discussed by the players during the experiment, notably their understanding, doubts, and strategies. This setup gave us access to the situated understanding of the players. It allowed us to verify if good performances in the game were linked to a valid understanding of influences and cartpole.

\begin{figure*}
  \centering
  \includegraphics[width=1\columnwidth]{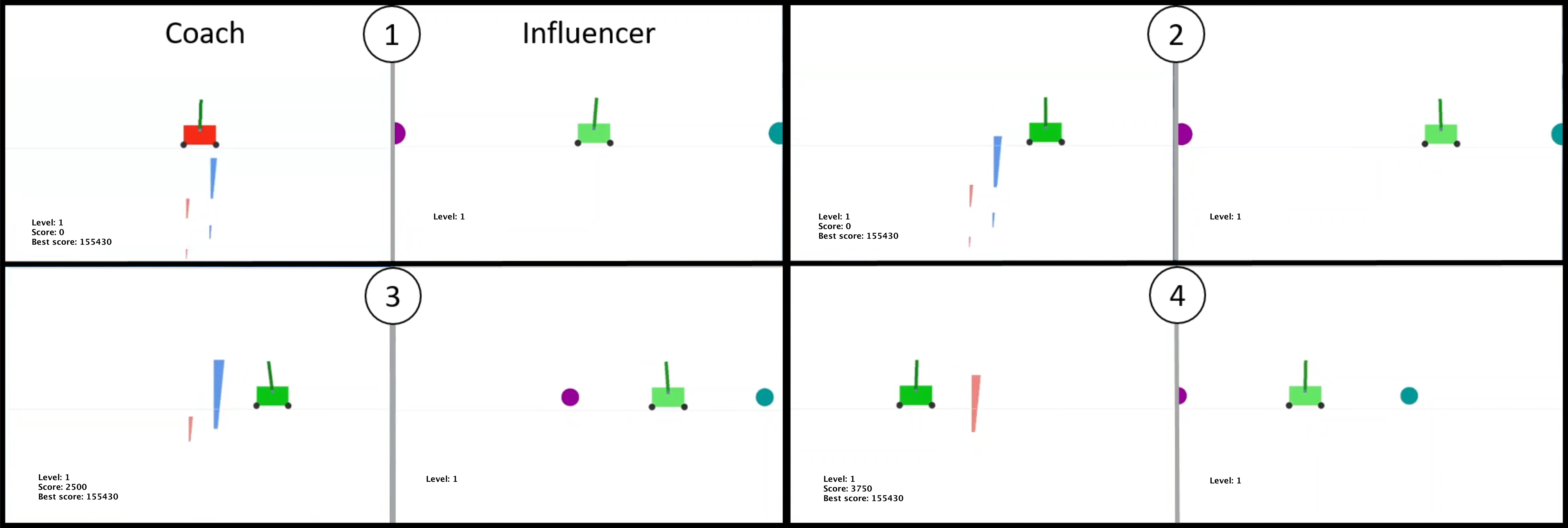}
  \caption{Screenshots from the session of two players (P31 and P32) during four instances in their cartpole game. Top left: the beginning of the game. It is Level 1 of the game and the coach can see the four gates (vertical blue and red lines) that need to be passed by the cartpole in their screen. On the coach's screen, there is also additional information such as score, level, and best score. The cart is red coloured on the coach's screen as it is not positioned in the correct part of the screen, which is to the right of the blue gate to pass the gate. On the influencer's screen, the player can see the influence circles, the cartpole, and the level of the game. The cartpole does not change colour on the influencer's screen. In instance 2, the cartpole has reached the right part of the screen and we can see on the coach's screen that the gate is passing, i.e. moving from the bottom to the top of the screen where it disappears. In instances 3 and 4, we see the positions of the influencer's circles which moves the cartpole to the correct part of the screen. The correctness of the position of cartpole, is made visible on the coach's screen with the green-coloured cartpole and the gate passing. In step 4, the coach sees that it is the end of the game as it is the last gate.}~\label{fig:screens}
    \vspace{-5pt}
\end{figure*}

In this two-player game, one player, \textit{the influencer}, can manipulate the influences and modify the behaviour of the cartpole. The other player, \textit{the coach} can see the goals to increase the score. In Figure \ref{fig:screens}, screenshots of four instances from a single game are represented to illustrate the elements in the respective screens of the coach and the influencer.

\begin{figure*}
  \centering
  \includegraphics[width=1\columnwidth]{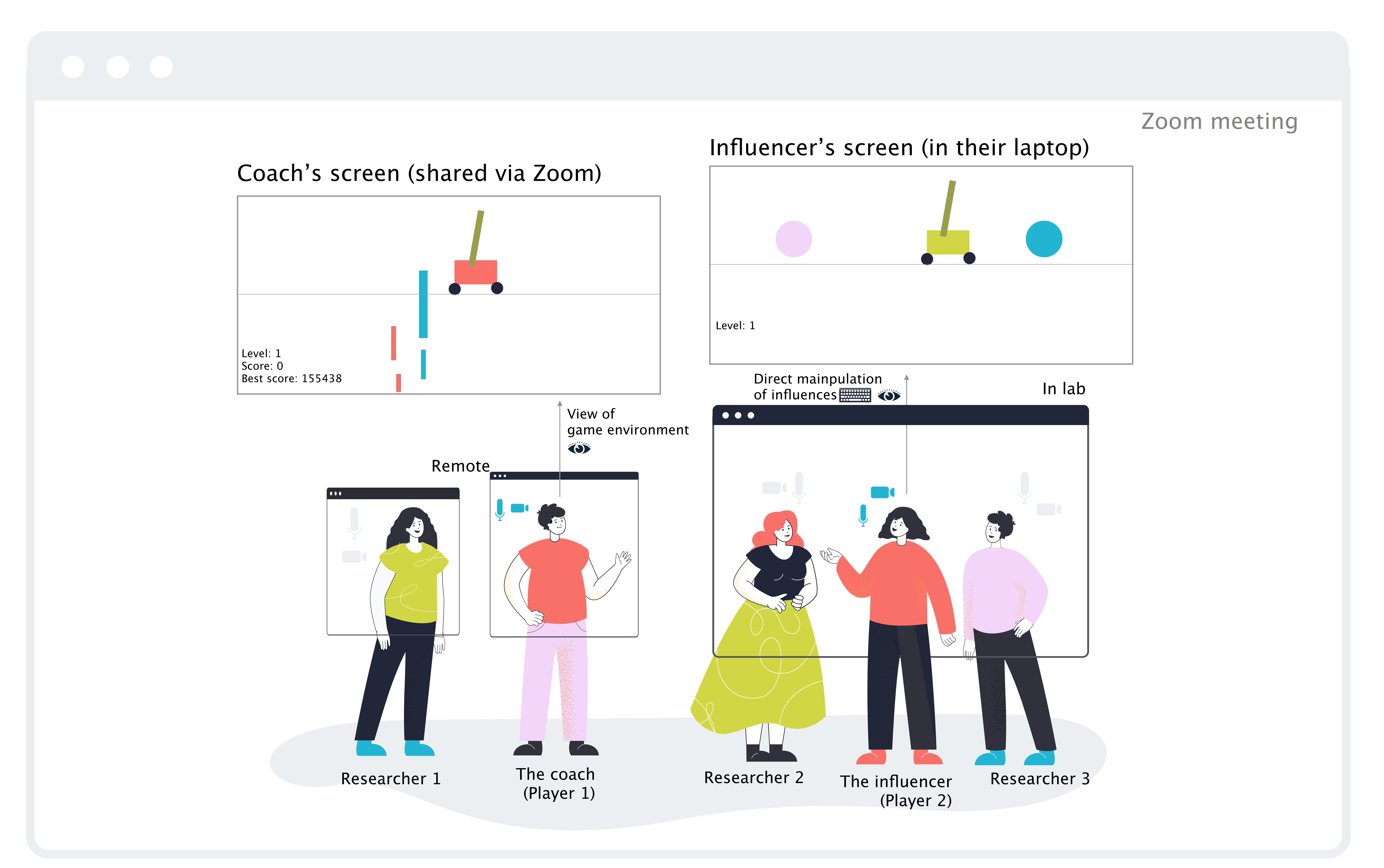}
  \caption{The figure illustrates the experimental setup which is a Zoom meeting. We see on the left part the remote participant with the coach's screen and the remote researcher. On the right side of the figure is illustrated the local participant with the influencer's screen and two researchers also present in the meeting room. }~\label{fig:setupexpe}\end{figure*}

The two players were playing together via a Zoom \cite{zoom}  video call. There were five people, three researchers, and two players, involved in the Zoom call, distributed over three places as illustrated in Figure \ref{fig:setupexpe}. The influencer was physically present with two researchers in a meeting room, playing on a laptop we had provided. The coach was playing remotely via Zoom. We shared the coach's gaming screen with them via Zoom screen share and this screen was hidden from the influencer's view, as was the influencer's screen from the coach's view. A third researcher was present remotely on the Zoom call.

Each session lasted around 45 minutes during which players performed the gaming experiment for 30 minutes. Before launching the Zoom meeting, the influencer was introduced to the control keys on the laptop's keyboard, which allowed them to modify the size and position of the influences on the screen. We explained the experimental setup to the influencer and they could try moving and resizing the influences by pressing the control keys a few times. Then we launched the Zoom video call, notified both participants that they would have to play the game together (i.e. not against each other), that both of them would see different elements on their screen, and that need to communicate with each other (think aloud) when playing the game. We further used the metaphor of ``hands and eyes'' to make sure that both players understood our briefing. The influencer was introduced as the hands, one who could act on the game, and the coach as the eyes, one who could see how to score. Half of the pairs (11/22) were notified that the cartpole was an autonomous agent with an AI-based model.

Both players were then asked to observe three \textit{hands-free} games. 
In those games, the influence circles were set to static medium-sized (Figure \ref{fig:sizenfluences}) to maintain the cartpole in the game as long as possible. After this observation, both players were asked to play the game between themselves for 30 minutes. During these 30 minutes, the players communicated with each other and tried their best to understand the game to jump levels. Players constantly talked with each other strategising while playing the game, and they were also allowed to \textit{pause} the game to plan with each other anytime during the game. All three researchers were silent observers during this gaming time. 



After the 30-minute gaming, we performed two short interviews. The first interview was an individual interview, where each participant was interviewed separately. The remote researcher took a separate Zoom call with the coach (remote participant) and the onsite researchers interviewed the influencer (colocated participant). This was a semi-structured interview, with questions about the player's understanding of the game, their strategies, if they felt to have an impact on the cartpole, about their collaboration with their co-player. We then ran a joint interview with the two participants and the three researchers, all together in the same Zoom meeting. In this second interview, which was the last part of our experiment, we showed the participants some screenshots of the coach's screen illustrating significant moments in their game such as when they lost control of the cartpole or when they succeeded to jump levels. Both participants jointly commented on their respective understandings of what had happened during those moments in the game. After this exercise with the screenshots, we brought up the fact that the cartpole in their game was controlled by an AI algorithm, this was new information for half of the participants (see Table\ref{tab:participants}), and questioned them further on their thoughts on the topic. For the participants who knew this before beginning the experiment, we also enquired whether they thought about the AI playing alongside them during their game. To all participants, as a concluding question, we asked if and how the existence of this AI algorithm affected their feeling of having an impact on the cartpole on using the given influences.


\subsubsection{Participants} 
We recruited 44 participants (22 male and 22 female), for our experiment through a myriad of channels such as handing out flyers and posting in many public spaces (such as cafes, restaurants, supermarkets, and small local shops), posting on social media, emailing local research communities, and by requesting the ones who signed up to play onsite to bring on a friend or family member to play together with them as a remote player. In the end, we had 10 related ones (friends or family) pairs and 12 strangers, making up the selection of 22 pairs of participants who played our cartpole game. All participants had native- or bilingual-level English speaking skills. Their demographic information is listed in table \ref{tab:participants}, along with expertise in Tech (from 1 to 4) and experience in gaming (from 1 to 4) which we collected in an initial screening survey. All participants received shopping vouchers worth \$30 as remuneration for their participation.

\begin{table}
    \centering
        \begin{tabular}{lll||c|cccc||c|cccc} 
          \hline
            \textbf{Team}& \textbf{Pair} & \textbf{AI} & \multicolumn{5}{c}{\textbf{Coach}} & \multicolumn{5}{c}{\textbf{Influencer}}  \\
           & & &{ID} &{Age} & {Gender} & {Tech} & {Game} &{ID} &{Age} & {Gender} & {Tech} & {Game} \\
            \hline
            \textbf{P1P2} & Strangers & No & P1 & 35-44 & F & 4 & 2 & P2 & 35-44 & F & 4 & 3 \\ 
            \textbf{P3P4} & Strangers & Yes & P3 & 18-24 & F & 4 & 2 & P4 & 35-44 & F & 2 & 1 \\ 
            \textbf{P5P6} & Related & No & P5 & 18-24 & F & 3 & 3 & P6 & 18-24 & F & 4 & 2 \\ 
            \textbf{P7P8} & Related & Yes & P7 & 25-34 & F & 4 & 3 & P8 & 35-44 & M & 3 & 4 \\ 
            \textbf{P9P10} & Strangers & No & P9 & 18-24 & M & 3 & 3 & P10 & 35-44 & M & 2 & 3 \\ 
            \textbf{P11P12} & Strangers & Yes & P11 & 18-24 & M & 4 & 4 & P12 & 25-34 & F & 4 & 2 \\ 
            \textbf{P13P14} & Strangers & No & P13 & 18-24 & F & 3 & 3 & P14 & 18-24 & F & 4 & 3 \\ 
            \textbf{P15P16} & Strangers & Yes & P15 & 25-34 & F & 3 & 2 & P16 & 25-34 & M & 3 & 2 \\ 
            \textbf{P17P18} & Related & No & P17 & 25-34 & F & 3 & 2 & P18 & 25-34 & M & 4 & 3 \\ 
            \textbf{P19P20} & Related & Yes & P19 & 18-24 & M & 3 & 3 & P20 & 18-24 & F & 2 & 2 \\ 
            \textbf{P21P22} & Strangers & No & P21  & 18-24 & F & 3 & 3 & P22 & 25-34 & F & 2 & 1 \\ 
            \textbf{P23P24} & Related & Yes & P23 & 18-24 & F & 3 & 2 & P24 & 18-24 & M & 4 & 3 \\ 
            \textbf{P25P26} & Strangers & No & P25 & 18-24 & F & 3 & 3 & P26 & 25-34 & M & 3 & 4 \\ 
            \textbf{P27P28} & Strangers & Yes & P27 & 25-34 & M & 3 & 3 & P28 & 35-44 & M & 3 & 2 \\ 
            \textbf{P29P30} & Related & No & P29 & 25-34 & F & 3 & 2 & P30 & 25-34 & F & 4 & 4 \\ 
            \textbf{P31P32} & Strangers & Yes & P31 & 18-24 & M & 4 & 4 & P32 & 25-34 & M & 4 & 4 \\ 
            \textbf{P33P34} & Related & No & P33 & 25-34 & M & 3 & 2 & P34 & 18-24 & M & 3 & 3 \\ 
            \textbf{P35P36} & Related & Yes & P35 & 35-44 & M & 4 & 4 & P36 & 35-44 & M & 4 & 3 \\ 
            \textbf{P37P38} & Related & No & P37 & 35-44 & F & 3 & 1 & P38 & 25-34 & M & 3 & 4 \\ 
            \textbf{P39P40} & Strangers & Yes & P39 & 18-24 & M & 3 & 3 & P40 & 25-34 & M & 3 & 4 \\ 
            \textbf{P41P42} & Related & No & P41 & 25-34 & F & 4 & 2 & P42 & 25-34 & F & 4 & 4 \\ 
            \textbf{P43P44} & Strangers & Yes & P43 & 25-34 & M & 4 & 4 & P43 & 18-24 & M & 3 & 3 \\

           \hline
        \end{tabular}
        
     \caption{The demographic details of the 22 teams of participants who played our cartpole game. Each team's ID is a combination of the two players' IDs. We also indicate if they were related (friends or family) or strangers and if they were told before the game that the cartpole is an AI system. For each participant we indicate, age, gender, tech savviness (1=not at all to 5=extremely good), and gaming experience (1=none at all to 5=extreme gamer).}
    \label{tab:participants}

\end{table}

\subsubsection{Data Collection} 
During the game, we recorded the screens of both participants. We also audio and video recorded the Zoom meetings from all sessions, consisting of the conversation between the participants and the two sets of post-gaming interviews. We collected the logs of the game that includes the position of the cart, the position of the gates, the position and size of the influences, the time, the number of steps of the cartpole's movements, and the action of the cartpole (whether it followed the model, a random action, or the influences). We faced a few minor issues in data collection with unexpected bugs and lack of memory space which resulted in few data gaps for four teams. We do not have the screen of the coach for P11P12 and P19P20. We also lack the screen of the influencer for P25P26 and the logs of the game for P41P42. These did not prevent us analysing the activity for all the teams, as we had the entire session recordings in Zoom. But, as a consequence, we cannot include the data of P41P42 in the quantitative analysis of our findings. The data collected during this experiment is handled under GDPR compliance.

\subsection{Data analysis} 
We performed both quantitative and qualitative analysis of the data collected in the experiment. In the first phase of our analysis, the qualitative data from the Zoom video recordings of the gaming and the subsequent player interviews were transcribed. In parallel, we compiled a database with all the logs collected from the game and we explored this data through descriptive statistics to figure out how teams performed in the game and progressed in mastering the use of influences toward the desired goal. 

Two researchers open-coded the transcripts of the game to perform an inductive thematic analysis \cite{thematicAnalysis}. From the familiarisation with the data emerged initial codes related to goals of the game - cartpole, gates, fail, score, success, the object of influences, circles, position, movements and effects, the instructions (go on the right or left), strategies of how to play the game and communication issues between player. These codes were then shared with the team members for asynchronous discussions on them. They were then refined to answer the questions of how the communication during the game informed us on the performance of the team. Following this direction, four themes emerged from four iterations of coding phases: Description of the scene, Instructions, Explanations and Strategies. A consensus was reached on the underlying sub-themes on discussing how participants could make sense of (1) the rules of the game and (2) the cartpole's behaviour or how they described the effects of the influences on the cartpole. The complete thematic analysis took place within six meetings separated by individual coding sessions. We also looked for quantitative measures of how the cartpole was impacted by the influences, and how much the teams could consecutively pass gates. Finally, we analysed the evolution of the performances, to understand whether experiencing the influences, could the influencers progressively make sense of their functioning, and use them effectively to control the cartpole. We juxtapose this mixed-method analysis to detail our findings in section \ref{sec:findings}.


\section{Prototype: Cartpole with Influences}
\label{sec:prototype}
\begin{figure}
  \centering
  \includegraphics[width=0.7\columnwidth]{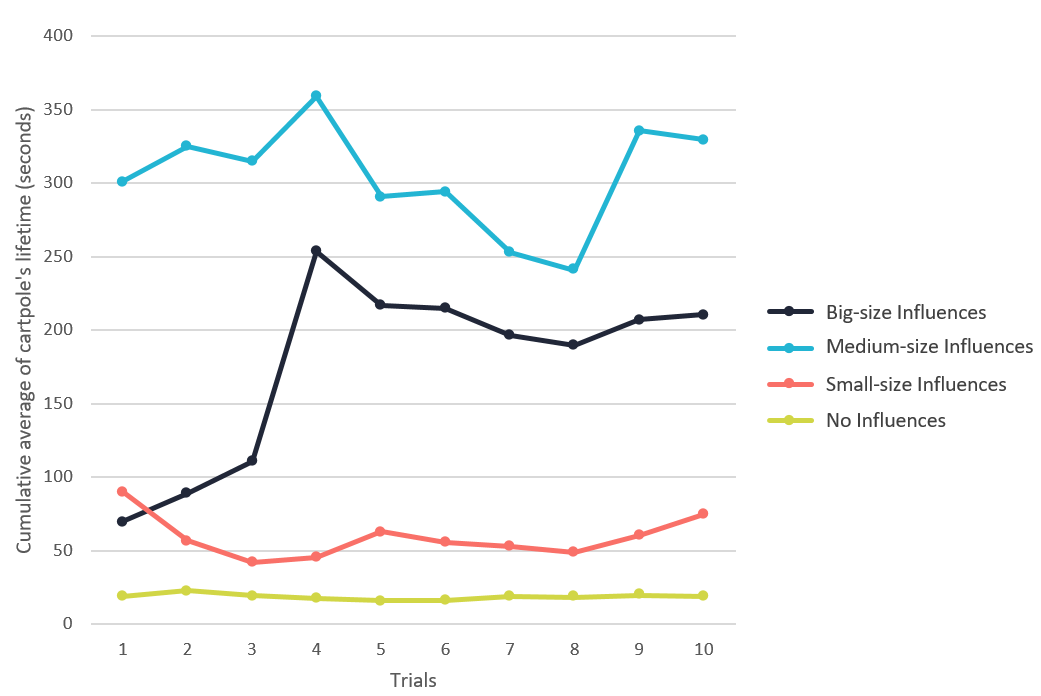}
  \caption{Cumulative average of cartpole's lifetime over ten trials under four conditions: No Influences, Small-size Influences, Medium-size Influences and Big-size Influences. The value in trial 1 corresponds to the cartpole's lifetime (in seconds) for the first trial. The value in trial 4 corresponds to the average lifetime of the cartpole over the trials 1, 2, 3 and 4.}~\label{fig:lifetime_influence}
\end{figure}

In our experiment, the policy of the model to keep the pole from falling has been implemented with the library \textit{d3rlpy.DiscreteSAC} through 710000 Reinforcement Learning (RL) steps \cite{OpenAIGym}. The model makes the cart to move from right to left in the screen to balance the pole to be erect upright on the cart. The model has not been fully trained, this is why it does not know the limits of the screen and can thus make the mistake of exiting the screen. The average lifetime of the game (in which the pole does not fall off the cart nor the cart exit the screen), calculated over 10 trials stabilises around 17 seconds, given a stochastic noise over actions of 0.1 (Figure \ref{fig:lifetime_influence}). 



\begin{figure}
\includegraphics[width=0.7\columnwidth]{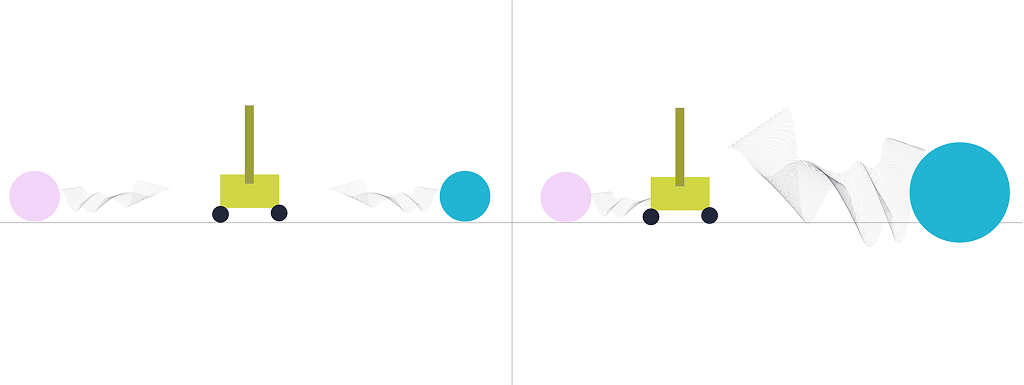}
\caption{Illustration of the impact of influences on the cartpole movement on two screens. On the left screen, the cartpole is stabilized with two influences of the same size. Both influences send equal "forces" that maintain the cartpole in the center of the screen. On the right screen, the cartpole is pushed to the left because of the bigger right influence}.~\label{fig:impactinfluences}
\end{figure}

To modify the action of the cartpole, we provide two \textit{influences}, circles in figure \ref{fig:impactinfluences}. Influences are a human-in-the-loop technique for training a reinforcement learning model \cite{JacquinPerezBoulard}. Humans can tune the agent's actions using influences, and then re-train the model including the modifications from the influences. We leverage this behind-the-scenes technique, by associating influences with interaction elements in a gaming environment which facilitates real time learning (fluid intelligence \cite{fluidIntel}). Influences, in the form of two circles in the interface, can be use for gaming with our simulated embedded AI agent, which is a cartpole. 

\begin{figure*}
  \centering
  \includegraphics[width=0.24\columnwidth]{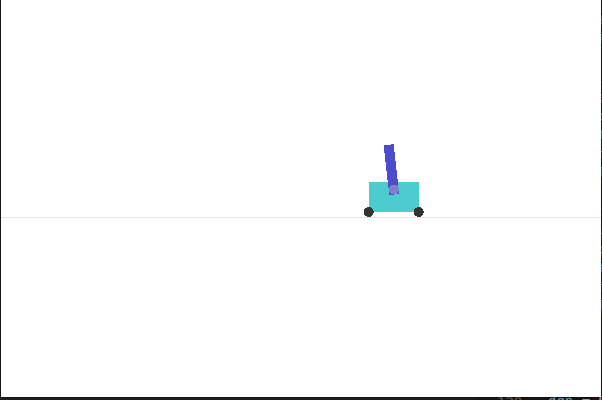}
  \includegraphics[width=0.24\columnwidth]{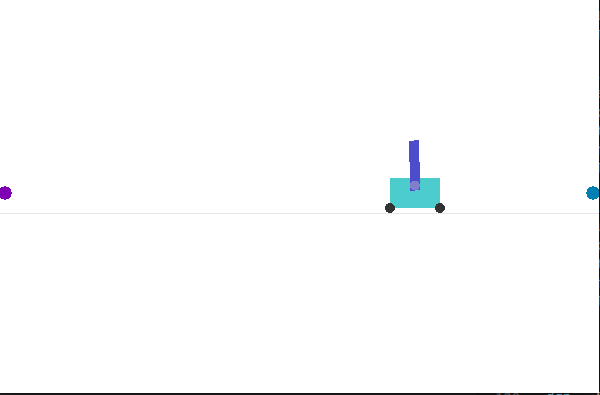}
  \includegraphics[width=0.24\columnwidth]{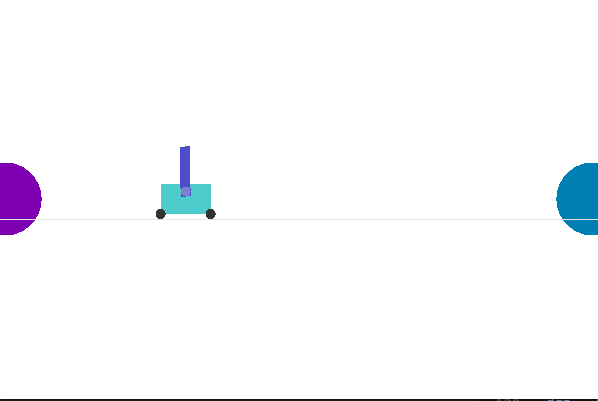}
  \includegraphics[width=0.24\columnwidth]{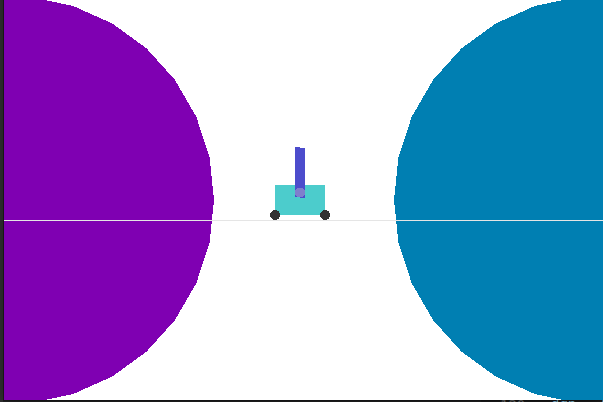}
  \caption{Screenshots from playing the game under static influences. When no influence is added (on left), the cartpole applies the learned model under a 10\% stochastic action. The trained model acts in the same stochastic configuration under small-size influences (center-left), medium-size influences (on center-right), and big-size influences (on right).}~\label{fig:sizenfluences}
\end{figure*}

The two added circles represent attracting influences on each extremity of the playground (i.e. the player's screen). The bigger the circles are, the stronger their influence is to attract the cartpole (see Figure \ref{fig:sizenfluences} and Figure \ref{fig:impactinfluences}). Similarly, the closer to the cart the circles are placed, the stronger they attract it. When the cart is close enough to a single influence, it will accelerate in its direction, this acceleration makes the pole fall in the opposite direction. Then the model will try to prevent the falling of the pole by moving the cart back under the pole, i.e. away from the closest influence to stabilise the pole.

To illustrate the effect influences have on the cartpole, we illustrate the evolution of the cartpole's lifetime in Figure \ref{fig:lifetime_influence}.
We define cartpole's lifetime as the time in seconds in which the cartpole remains on the screen, with the pole upright on the cart.
The four lines in the line graph (Figure \ref{fig:lifetime_influence}), illustrate the evolution of cartpole's lifetime on sustaining no, small, medium, and big-sized influences (with set of influences illustrated in Figure \ref{fig:sizenfluences}). We observed medium influences to lengthen the lifetime, more than any other combination of influences. More qualitatively, on having no influences or small-sized influences, the game always ended with the cartpole exiting the screen. On having big-sized influences, the game always ended with a fall. With medium-sized influences, the game ended with the cartpole exiting the screen in 6 trials out of 10. The explanation for these behaviours is that: small-sized influences are never strong enough to unbalance the pole when attracting the cart. Thus in the case of small-sized influences, the model does not need to correct any unbalance by sending the cartpole to the center of the screen. Whereas with big-sized influences, the model can never counter the unbalance of the pole created by the influences, thus ultimately the game ends with the pole falling off the cart.  

Algorithm \ref{alg:action} shows how the action of the cartpole is chosen between the trained model, the influences, or a random action, due to the stochastic noise. Algorithm \ref{alg:action} uses the $inf$ vector whose computation is later detailed in algorithm \ref{alg:infvec}.

\begin{algorithm}
\caption{Decision of cartpole's next action}
\label{alg:action}
\begin{algorithmic}
\Statex $action$ = vector of the action to execute;
\Statex $m$ = vector of model's preference;
\Statex $inf$ = vector of influences' preference;
\Statex
\For{each step}
    \If{stochastic}
        \State $action \gets random$
    \Else
        \State $action \gets maximum$ $vector$ $in$ [$m * inf$]
\EndIf
\EndFor
\end{algorithmic}
\end{algorithm}

Algorithm \ref{alg:infvec} details how the vector $inf$, the vector of influence preference, is computed from a set of influences. $inf$ is computed in two steps, one step for each influence, namely the right and the left. For each influence, the vector $res$ computes a preference vector that takes into account the distance between the influence and the cartpole (in $location\_related\_weight$), the speed and direction of the cart with respect to the influence (in $speed\_related\_weight$) and the size of the influences (in $intensity$).
Then both influence vectors are summed in the $inf$, which is then used in the algorithm \ref{alg:action}.

\begin{algorithm}
\caption{Computation of inf vector}\label{alg:infvec}
\begin{algorithmic}
\Statex $inf$ = vector of influences' preference;
\Statex $location\_related\_weight$ = strength of the influence in the current state and position;
\Statex $speed\_related\_weight$ = strength of the influence related to the current state and the speed of the cart;
\Statex $intensity$ = strength given to the influence by the user;
\Statex $res$ = preference related to the current state;
\Statex
\For{each influence}
    \State Calculate distance between the cart and the influence
    \State $location\_related\_weight \gets$ discriminated distance's value to have stronger impact when located closer to the cart 
    \State Convert the direction of the influence into numeric values in a $vector$
    \State $res \gets vector * location\_related\_weight$ 
    \State Calculate the $speed\_related\_weight$ according to inertia
    \State $res \gets res^{0.5} * speed\_related\_weight$
    \State $res \gets res * intensity$
    \State $inf \gets inf + res$ \Comment{Sum the weighted preference of each influence given the state}
\EndFor
\Statex
\Return $inf$
\end{algorithmic}
\end{algorithm}

\section{Findings}
\label{sec:findings}


\subsection{Analysis of the Experiment}

\begin{table}
    \centering
        \begin{tabular}{l|c|cccc|c|ccc} 
          \hline
            \textbf{Team}& \textbf{Time} &  \multicolumn{4}{c|}{\textbf{Gates}} &
            \textbf{Average of  \textit{ConsGates}} & \multicolumn{3}{c}{\textbf{Action of the cartpole}}  \\
           & & {Failed} &{Passed} & {Total} & {Ratio} & & {Influence} &{Model} &{Stochastic} \\
            \hline
            \textbf{P13P14} & 07:01 & 29 & 10 & 39 & 25.64\% & 0.34 & 9.70\% & 75.99\% & 14.31\% \\
            \textbf{P19P20} & 11:45 & 24 & 12 & 36 & 33.33\% & 0.50 & 9.82\% & 76.24\% & 13.94\%  \\
            \textbf{P39P40} & 19:21 & 38 & 26 & 64 & 40.63\% & 0.68 & 10.24\% & 75.77\% & 13.99\%  \\
            \textbf{P5P6} & 10:11 & 39 & 28 & 67 & 41.79\% & 0.72 & 10.43\% & 76.24\% & 13.33\%  \\
            \textbf{P1P2} & 16:15 & 31 & 24 & 55 & 43.64\% & 0.77 & 17.93\% & 68.57\% & 13.50\%  \\
            \textbf{P23P24} & 09:18 & 24 & 21 & 45 & 46.67\% & 0.84 & 7.67\% & 78.18\% & 14.15\%  \\
            \textbf{P33P34} & 17:37 & 32 & 27 & 59 & 45.76\% & 0.84 & 12.32\% & 74.26\% & 13.42\%  \\
            \rowcolor{lightgray} \textbf{P25P26} & 10:23 & 16 & 19 & 35 & 54.29\% & 1.19 & 7.07\% & 79.11\% & 13.82\%  \\
            \rowcolor{lightgray} \textbf{P21P22} & 07:50 & 23 & 32 & 55 & 58.18\% & 1.39 & 4.43\% & 81.70\% & 13.87\%  \\
            \rowcolor{lightgray} \textbf{P11P12} & 13:46 & 29 & 41 & 70 & 58.57\% & 1.41 & 7.68\% & 78.03\% & 14.29\%  \\
            \rowcolor{lightgray} \textbf{P43P44} & 15:34 & 21 & 30 & 51 & 58.82\% & 1.43 & 6.63\% & 79.65\% & 13.72\%  \\
            \rowcolor{lightgray} \textbf{P3P4} & 11:22 & 16 & 27 & 43 & 62.79\% & 1.69 & 13.19\% & 72.72\% & 14.09\%  \\
            \rowcolor{lightgray} \textbf{P29P30} & 14:44 & 14 & 33 & 47 & 70.21\% & 2.36 & 12.97\% & 73.09\% & 13.94\%  \\
            \rowcolor{gray} \textbf{P17P18} & 23:01 & 13 & 48 & 61 & 78.69\% & 3.69 & 8.86\% & 76.74\% & 14.40\%  \\
            \rowcolor{gray} \textbf{P15P16} & 24:39 & 14 & 64 & 78 & 82.05\% & 4.27 & 6.64\% & 79.31\% & 14.05\%  \\
            \rowcolor{gray} \textbf{P37P38} & 21:09 & 8 & 41 & 49 & 83.67\% & 4.56 & 6.09\% & 79.90\% & 14.01\%  \\
            \rowcolor{gray} \textbf{P35P36} & 22:01 & 12 & 57 & 69 & 82.61\% & 4.75 & 6.81\% & 79.29\% & 13.90\%  \\
            \rowcolor{gray} \textbf{P31P32} & 21:19 & 10 & 57 & 67 & 85.07\% & 5.18 & 15.16\% & 70.80\% & 14.04\%  \\
            \rowcolor{gray} \textbf{P27P28} & 24:14 & 17 & 103 & 120 & 85.83\% & 6.06 & 5.99\% & 79.89\% & 14.12\%  \\
            \rowcolor{gray} \textbf{PP7P8} & 10:24 & 3 & 30 & 33 & 90.91\% & 7.50 & 6.87\% & 79.23\% & 13.91\%  \\
            \rowcolor{gray} \textbf{P9P10} & 23:53 & 7 & 79 & 86 & 91.86\% & 11.29 & 4.81\% & 81.41\% & 13.78\%  \\

           \hline
        \end{tabular}
        
     \caption{This table indicates for each team: the time spent playing in minutes and seconds, the number of gates the teams failed to pass, succeeded to pass, and the total number of gates they encountered (which is the sum of failed and passed gates), and the success ratio (passed gates divided by the total number of gates). The average of \textit{ConsGate} is the average of the number of consecutively passed gates. The last three columns on the cartpole action show the percentage of steps the cart actioned following either the influences, the model, or stochastic. The table orders the teams based on the average of \textit{ConsGates} and indicates two thresholds: average of \textit{ConsGates} is less than 1 for low-level teams, between 1 and 3 for intermediate-level teams, and more than 3 for high-level teams indicating their performances}
    \label{tab:TablePart1}

\end{table}

\subsubsection{Performances of the teams}
\label{sec:performances}
To illustrate the performances of the teams, we had the choice to consider the final score, the achieved level, the number of successful games, or the number of gates. As our goal in this contribution is to show how non-experts can configure AI systems in the Actionable AI paradigm, we chose to focus on the number of gates each team was able to pass. We also considered the number of consecutive gates passed by the teams before the cartpole failed, either on having the pole fall off or on the cartpole exiting the playground. First it required to learn to maintain the pole on the cart, meaning to keep the angle between the pole and vertical axis below a threshold of 80°, and second to keep the cartpole in the screen. From this point, we call \textit{ConsGates} the number of consecutively passed gates a team could achieve. The more a team was able to pass consecutive gates, the higher the level of the team's performance was. Indeed this is an indicator that the team understood that they lose the game when the pole falls or when the cartpole exited the screen. It also indicates that the influencer had found ways to manipulate the influences efficiently toward the goal of the team. 


The total number of gates encountered varies from 33 gates for P7P8 to 120 gates for P27P28 (See Table \ref{tab:TablePart1}). This indicates the percentage of passed gates for each team, displaying a distinct variation in terms of performances with low-level teams such as P13P14 (26\%) or P19P20 (33\%) and high-level teams, P9P10 (92\%) or P7P8 (91\%). Finally, the average \textit{ConsGates} a team could pass varies between 0.34 for P13P14 and 11.29 for P9P10. Qualitatively, we were able to collect several quotes (see table \ref{tab:quoteGame}) during the game or during the interviews on how participants were able to make sense of the rules of the game and what would make them win or lose. 16 out of 22 coaches, were able to have a good enough understanding to guide their influencer during the experiment.

\begin{table}
    \centering
        \begin{tabular}{l|ccp{9cm}} 
          \hline
            \textbf{Participants} & \textbf{Role} & \textbf{Source} & \textbf{Verbatim}
            \\ \hline
            \textbf{P6} & Influencer & During the game & To win it must remain stable. \\ \hline 
            \textbf{P7} & Coach & Individual interview & The vehicle has to be in the center if I want the bar [gates] going up. \\ \hline 
            \textbf{P17} & Coach & Individual interview & we were getting scores once the card was shifting right and left and it was hitting the lines [gates]. \\ \hline 
            \textbf{P24} & Influencer & During the game & The point of this game is to keep the car in the middle, as long as possible \\ 
            \textbf{P23} & Coach & During the game & We should also keep the stick the green stick in the middle in equilibrium. \\ \hline 
            \textbf{P25} & Coach & Individual interview & I think we had to keep the balance of the green stuff on the car, have the car moving throughout the screen, pass the little things [gates] coming from below, when passing the thing and the thing was moving up, I think it was when we had the points. \\ \hline
            \textbf{P35} & Coach & During the game & Okay, so when the cart gets off the screen, we lose apparently. \\ \hline 
            \textbf{P42} & Influencer & Individual interview & There is a cart moving by itself and I have to influence its movements by changing the size and the position of two bubbles and I have to keep the green bar on top of the cart balanced if it falls we lose and if the cart goes off-screen we lose too an we have to collect things that I can’t see. \\

           \hline
        \end{tabular}
        
     \caption{Quotes of participants illustrating their understanding of the rules of the cartpole game}
    \label{tab:quoteGame}

\end{table}

\subsubsection{Time played}
The time each team spent playing and interacting with the cartpole game is indicated in table \ref{tab:TablePart1}. Even if all teams had exactly 30 minutes to play the game interacting with the cartpole, they could also pause the game. This explains the broad diversity in time spent on the game itself, between 7 minutes for P13P14 and almost 25 minutes for P15P16. During these pauses, teams usually took time to discuss their understanding of the game, elicit strategies, and coordinate themselves. Some teams needed to talk more than others and it resulted in having teams that had little time to play.

\subsubsection{Influence on the cartpole's behavior}
During the experiment, by using the influences, participants affected the behaviour of the cartpole.
In table \ref{tab:TablePart1}, we see the actions of the cartpole, for each team, how much the cartpole's behaviour followed the model, a random action, or the influence given by the influencer (controlling the two circles on their screen using the keyboard). Here the influence is accounted for only when the cartpole moves in opposition to the model. Indeed, when the autonomous cartpole follows the model action, when the influences are not strong enough to compete with the model. In this case, there are two possible situations, either influences are positioned very far from the cartpole and have almost no effect on it, or they are close and strong enough but the model is even stronger. Another reason to account for the action of the cartpole is that the influences and model are in unison, they indicate to follow the same direction, and in that case, this is also recorded as the model action.

In table \ref{tab:TablePart1}, we observe that the number of times the cartpole follows the influences may vary from 5 \% to 17 \%.
There is no correlation between this number and the performance achieved by the team.
For instance teams P9P10 and P27P28 showed a very good level of performance and did not rely strongly on influences.
On another hand, P31P32 also had a high score while strongly relying on influences.
Similarly, there are teams with low levels of performance such as P23P24 with little influences (8 \%) or P1P2 putting a lot of influence (18 \%) (further analysed in section \ref{sec:Strategies}).

\subsubsection{Three levels of performances: low, intermediate and high}

We analysed performance of teams and created three levels according to the average of \textit{ConsGates} of the teams. 
\begin{itemize}
    \item Low-level teams with an average of \textit{ConsGates} less than 1 \item Intermediate-level teams with an average of \textit{ConsGates} between 1 and 3
    \item High-level teams with an average of \textit{ConsGates} higher than 3
\end{itemize}

This is denoted in table \ref{tab:TablePart1} using higher levels of grey background to indicate higher-level teams. As a way to confirm that \textit{ConsGates} is the correct indicator of performance, we see that low-level teams have a ratio of passed gates over a total number of gates below 50\%, for intermediate-level teams the ratio is between 50\% and 75\%, and for high-level teams it is above 75\%.

For low-level teams, the average of \textit{ConsGates} being less than 1  means that the teams of this group were not able to pass more than one gate consecutively. It shows that influencers could not make sense of how to stabilise the cartpole and keep it on the screen without falling the pole. The intermediate-level teams were able to pass on average between one and two consecutive gates. This suggests their ability to let the cartpole go on one side of the screen and then come back on the other side. It is a good indicator that the influencers mastered the influences enough to keep the pole in the center of the screen. The high-level teams, on average passed more than three consecutive gates. Some teams even reached 6 (P27P28), 7 (P7P8), or 11 (P9P10) average of \textit{ConsGates}. It demonstrates their adeptness in utilizing influences to navigate gates effortlessly.

In sum, among the 22 teams involved in the experiment, 14 teams showed an ability to use influences to modify the cartpole's behaviour towards a specific goal. We did not observe any impact of the type of pairs (strangers or related), on the performances of the team. Similarly, no differences in performances were observed based on delivering the information that an AI algorithm was controlling the cartpole, pre- or post-game. 


\subsection{Incremental understanding of the influences and the cartpole}
\label{sec:understanding}


\subsubsection{Progression in consecutive gates passed}
\begin{figure}
  \centering
  \includegraphics[width=0.6\columnwidth]{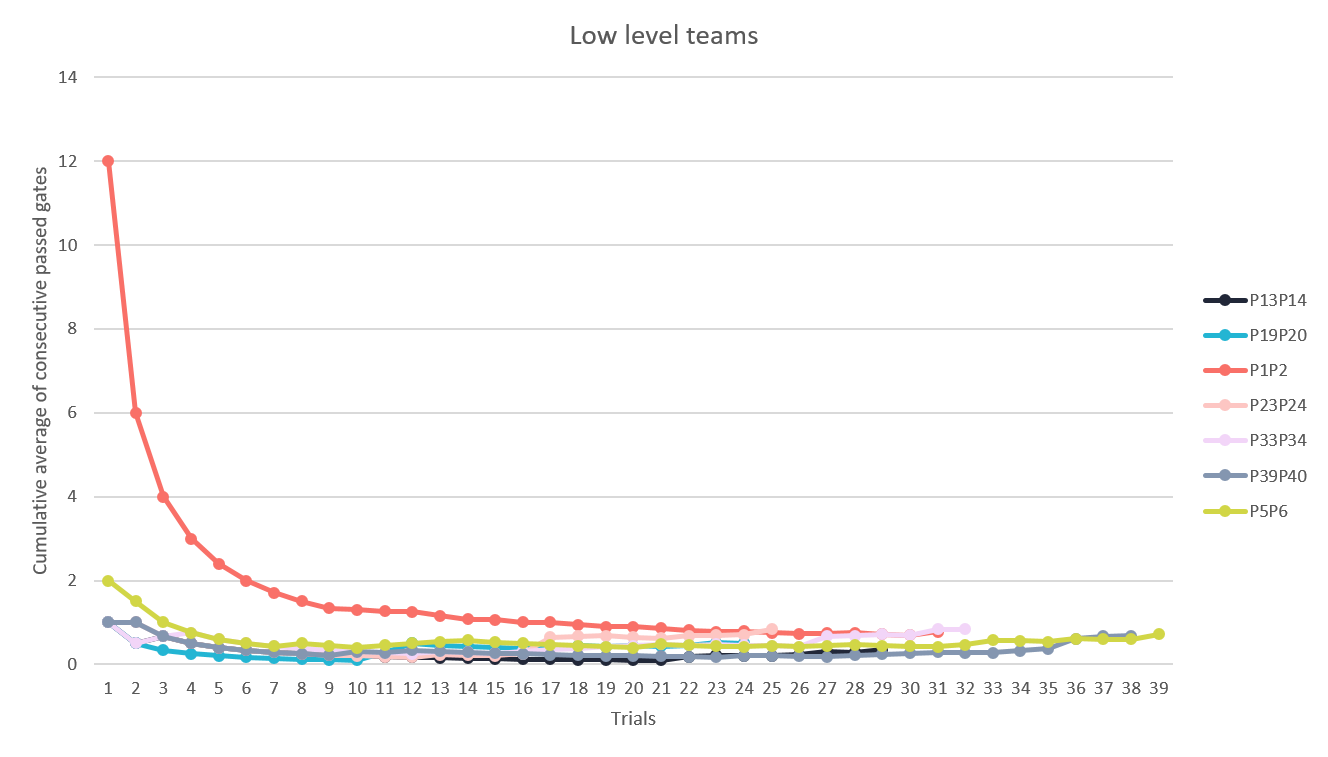}
  \includegraphics[width=0.6\columnwidth]{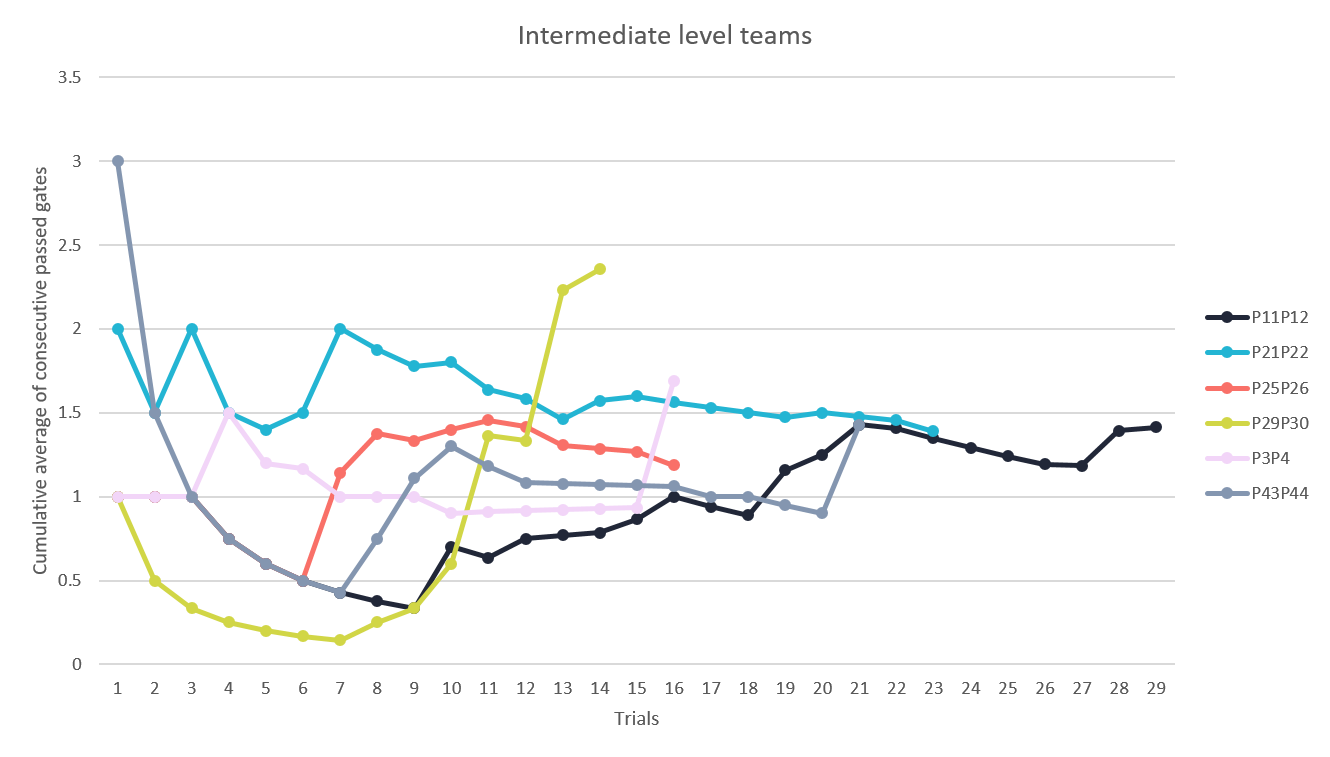}
  \includegraphics[width=0.6\columnwidth]{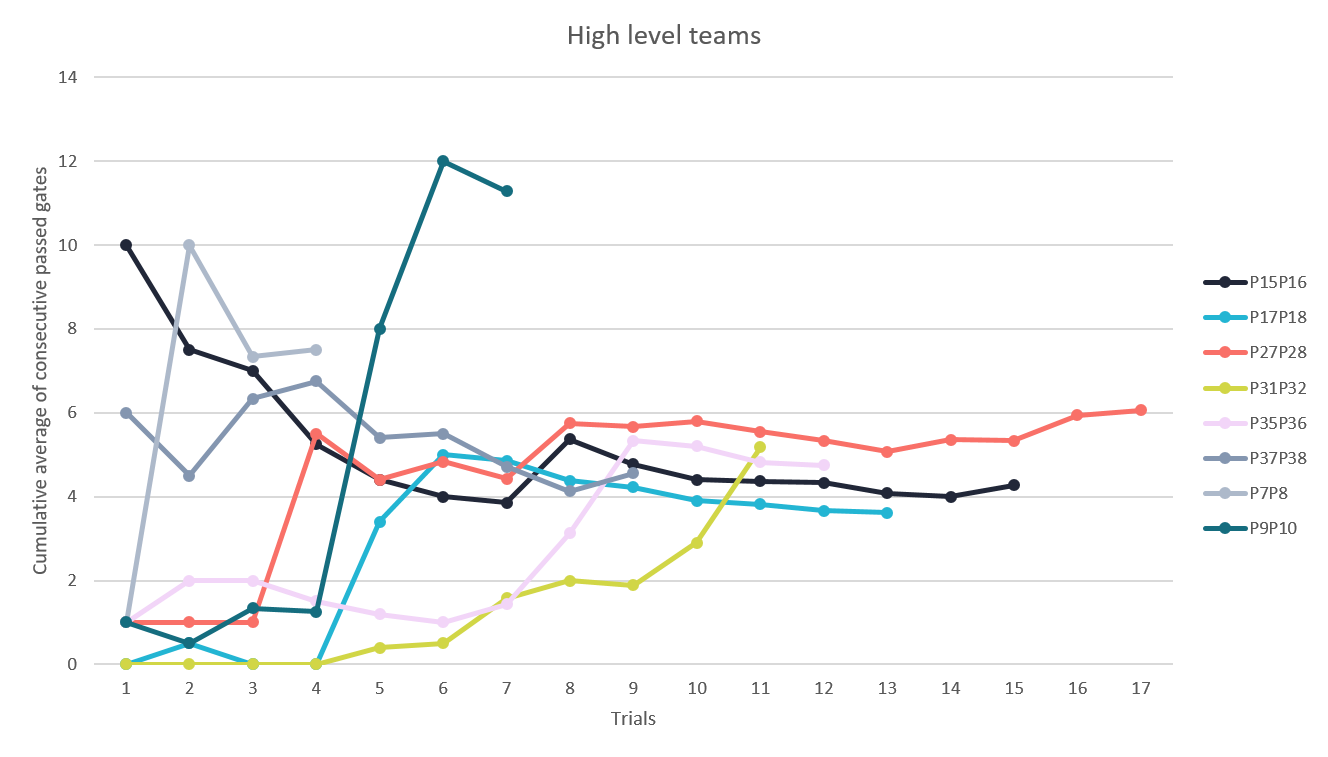}
  \caption{This figure represents on three graphs, three different levels of teams, based on their performance. It shows the cumulative average of \textit{ConsGates} over their trials in the experiment. By denoting it a trial, we define the phase of the game between two fails of the cartpole. In each trial, we looked at the number of gates passed and indicated the number until the next trial. If no gates were passed for the trial, we indicated 0. For instance, the value of the first trial is the exact number of gates passed in the first trial. The value of the third trial is the average of \textit{ConsGates} in the first, second, and third trials. The last trial of each team is the average \textit{ConsGates} for that team, also mentioned in the table \ref{tab:TablePart1}. }~\label{fig:3GroupsTeams}
\end{figure}

To illustrate how teams progressed in the experiment, we further analysed the number of consecutive passed gates (\textit{ConsGates}). We represent the cumulative average of \textit{ConsGates} for each group i.e. low-level teams, intermediate-level teams, and high-level teams in Figure \ref{fig:3GroupsTeams}. In the first graph of \ref{fig:3GroupsTeams}, we see the evolution of the average of \textit{ConsGates} for low-level teams. All teams in this low-level group remain between 0 and 1, and in comparison to the two other groups, they have an important number of trials between 24 (P19P20) and 39 (P5P6). In P1P2's first trial, they succeeded to pass 12 consecutive gates. During this first trial, they did not change the influences which were set as medium-size (Figure \ref{fig:sizenfluences}) during the hands-free mode. After this initial very-good performance, they never succeeded to pass more than one gate except in their last trial (See trial no.31 in Figure \ref{fig:3GroupsTeams}) where they passed 3 consecutive gates. These results suggest that P1P2 started to experience the use of influences late in their game (after 12 passed gates) and luckily achieved an initial high performance. We note in addition that all the low-level teams got a better first trial than the following ones as they also benefited from the hands-free setting of influences. We can emphasise that even if the average of \textit{ConsGate} of low-level teams remains under 1, we observe a tendency to improve with the number of trials.

The second graph of figure \ref{fig:3GroupsTeams} represents the same information (cumulative average of \textit{ConsGates}) for intermediate-level teams. There are fewer trials for this group when compared with low-level teams, between 14 trials (P29P30) and 29 (P11P12) trials, and all teams in the groups succeed to pass between 1 and 3 gates consecutively. We note for all teams except for P21P22 and P25P26, that their last average \textit{ConsGates} is their best score. For P21P22 and P25P26 even if the final score is not their best one, their average of \textit{ConsGates} does not drop as we had previously observed for P1P2 in the low-level groups graph. These results suggest that all intermediate-level teams had grasped how to make the cartpole oscillate from left to right using influences to continuously pass consecutive gates.

The third graph of figure \ref{fig:3GroupsTeams} shows the cumulative average of \textit{ConsGates} for the high-level teams. The number of trials is again lower than for the intermediate teams ranging from 4 trials (P7P8) to 17 trials (P27P28). There is a wider variation in the evolution of the average for this group. For the two groups intermediate-level and high-level, we see a learning phase. At the very beginning, the teams usually kept the position of the influences very same they were set in the experiment's ``hands-free'' mode. When starting to interact with the influences,  participants explored different positions of influences and performances dropped during this learning curve. This exploration resulted in the elaboration of an understanding of how the influences impacted the cartpole which grounded the improvements in performances. This finding is correlated with the time spent experiencing the cartpole. On average, it is around 13 minutes and 12 minutes respectively for low-level and intermediate-level teams, while it is around 21 minutes for high-level teams.

\subsubsection{Phases of the understanding}

 \begin{figure}
  \centering
  \includegraphics[width=1\columnwidth]{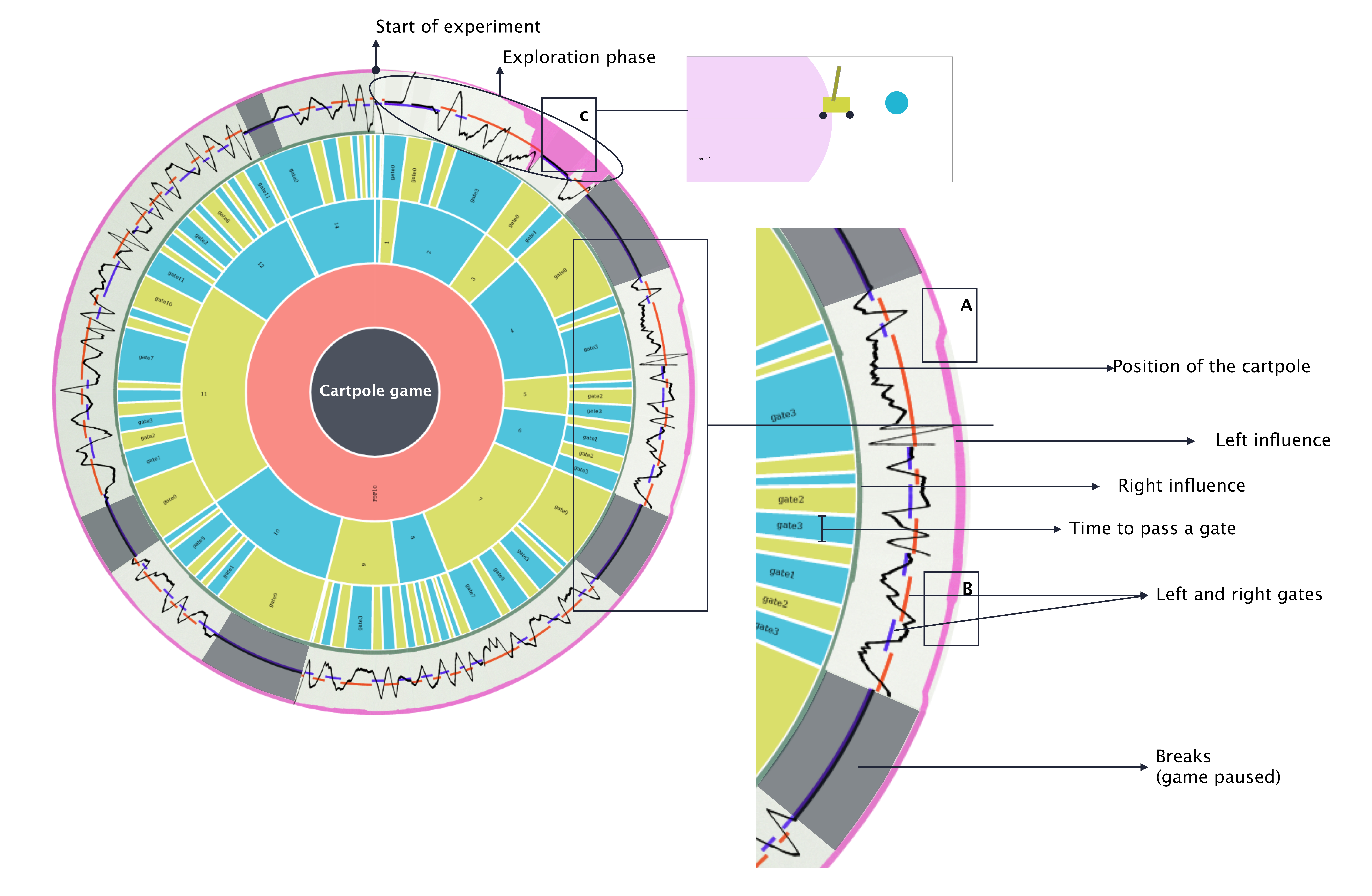}
  \caption{This visualisation shows the complete experiment of P9 and P10, who was the best performing team in our experiment, exhibiting a working knowledge of the cartpole acquired via their exploration phase with the influences, and succeeding in configuring it to achieve their goals in the game. }~\label{fig:P9P10}
\end{figure}

Along this evolution, influencers explored different strategies in the use of influences and were able to develop an understanding of how they could use the influences to improve their performances. We introduce in figure \ref{fig:P9P10} a visualisation that captures the 30-minute experiment of P9P10, visualised using D3.js \cite{D3}. The full circle on the left side of the figure represents their entire cartpole game. The start of the experiment is marked on the top and the dark-grey blocks indicate their breaks where the experiment was paused. In the sliced sequence of the game presented on the right part of figure \ref{fig:P9P10}, the light-grey external ring of the circle represents the screen. At each step of the game, the position of the cartpole is denoted with a black route. The size and positions of influences (left influence in purple and right influence in grey) and the position of the gates (blue and red lines) are also illustrated within the light-grey screen. In the rectangle marked as A in the figure, the influencer moved the left influence to the centre of the screen. As a consequence, the cartpole is pushed to the right of the screen and takes an extended amount of time to pass the red gate. In the rectangle marked as B in the figure, the size of the left influence (purple) is decreased. This visualises the time spent for a team to pass a gate.

In the full circle on the left side of figure \ref{fig:P9P10}, we can see the beginning of an exploration phase. P9P10 start their game by using really small-sized influences. This results in two consecutive exits of the cartpole, the first on the right, the second on the left. Then they increase the sizes of the influences a little and that keeps the cartpole in the centre of the screen. The rectangle marked as C in the figure shows moments where P10 (the influencer) had increased the left influence (in purple) a lot. This was the only time where P10 had made one influence big at this size and what P10 could see on their screen is illustrated alongside the rectangle marked as C in figure \ref{fig:P9P10}. This representation allowed us analyse extensively the path each team went through during the experiment.

 \begin{figure}
  \centering
  \includegraphics[width=1\columnwidth]{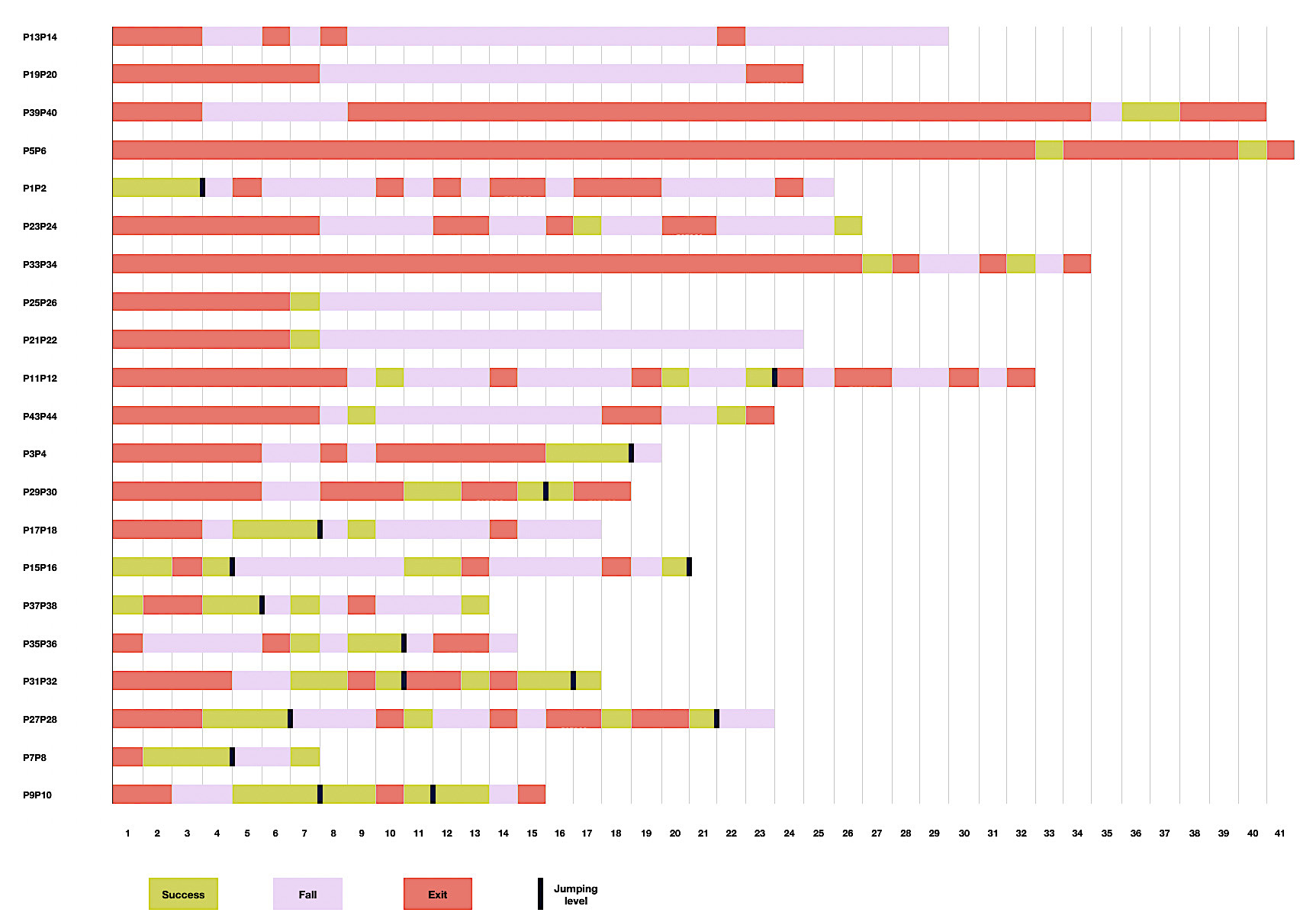}
  \caption{This graph illustrates for each team how they performed in each game they started. The outcome is either a success (the team succeeded to pass all the gates), a fall (the pole fell-down) or an exit (the cartpole exited the screen) The black lines show the change of game level.}~\label{fig:successfallexit}
\end{figure}

In a more condensed view, we present in figure \ref{fig:successfallexit} the sequence of success, falls and exits for each team during their game. In the learning curve, we observe that the high-level teams faced the two types of fails before progressing and gaining some successes. For low-level teams, we often see a long sequence of the same type of fail. P5P6 and P39P40 faced a high number of exits while P19P20 and P13P14 let the pole fall much more. In this context, it is really difficult for influencers to make sense of the impact of influences. Indeed, during the experiment, they were testing various sets of influences to understand their impact on the cartpole:

\begin {itemize}

\item Both influences on the same side of the cartpole,
\item Influences overlap on each other,
\item Extremely small influences that were almost not visible.

\end{itemize}

Unfortunately for them, those set of influences were inefficient in directing the cartpole movements. It delayed their understanding of the functioning of influences. For intermediate-level teams, we observe a progression, with a first phase similar to low-level teams and then an alternation of successes, falls and exits. The alternation is the one that allows influencers to understand how influences work.

The dark lines in the sequence of games outcome indicate the change in level. If we look at the 63 successful games that were followed by another game, they are followed by a success in 25 cases (40\%), by a fall in 20 cases (32\%) and by an exit in 18 cases (28\%). We observed that for the 15 successes that resulted in a new higher level, 9 were followed by a fall, 2 by an exit, and 4 by a success. It confirms the need for adaptation in the new level due to the disturbances (see Table \ref{tab:level}). This is further detailed in the next section.

\subsubsection{Description of how the influences work}

Influencers relied on several terms to describe the effect of influences, such as \textit{magnets, attractors, wind, repulsive magnets, notion of gravity,} and \textit{anti-gravity}. During the experiment, participants shared with their co-player how they made sense of the game. We could also collect during the interviews how people understood the influences' impact on the cartpole and more widely the purpose of the game. Table \ref{tab:quoteDoubts} lists quotes that illustrate doubts and difficulties influencers experienced during the game and how they shared it with their co-player. These results suggest the perception of ML model controlling the cartpole by the players, who did not get the same result on repeating their actions, and who understood that the cartpole was also moving autonomously even during their inaction.

\begin{table}
    \centering
        \begin{tabular}{l|ccp{9cm}} 
          \hline
            \textbf{Participants} & \textbf{Role} & \textbf{Source} & \textbf{Verbatim}
            \\ \hline
            \textbf{P2} & Influencer & During the game & I am not sure I understand, too responsive \\ \hline 
            \textbf{P6} & Influencer & During the game & I have one problem, I don't understand. \\ \hline 
            \textbf{P10} & Influencer & During the game & No more pressing any key anymore, I'm not sure I have some effects \\ \hline 
            \textbf{P16} & Influencer & During the game & I don’t correlate what I’m doing with the circles. \\ \hline 
            \textbf{P18} & Influencer & During the game & it's not me I don't control this thing [cartpole] really \\ \hline
            \textbf{P20} & Influencer & During the game & I didn't figure out yet how is really working I think it's something with balance but i'm not sure. \\ \hline 
            \textbf{P28} & Influencer & During the game & I don’t think I control anything. \\ \hline 
            \textbf{P34} & Influencer & During the game & the car is moving by itself there's some factor I don't know what it is. \\ \hline
            \textbf{P42} & Influencer & During the game &I think I don't know how to control the cart,  I don't know how to control it at all. \\ 
            \textbf{P41} & Coach & During the game &Right now you're not controlling it? it's just moving on its own ? \\
            \textbf{P42} & Influencer & During the game &it’s moving by itself, I think I can influence its movements. \\

           \hline
        \end{tabular}
        
     \caption{Quotes of participants illustrating the difficulties and doubts influencers experienced}
    \label{tab:quoteDoubts}

\end{table}

In table \ref{tab:quoteInterpretation} we list quotes that illustrate how participants hypothesised the use of influences, their intuitions, and interpretations. Table \ref{tab:quoteRepresentations} lists quotes where influencers described the impact of influences on the cartpole behavior. These results showcase the operational understanding of the cartpole gained by the players in managing the influences and observing their impact on the cartpole.

\begin{table}
    \centering
        \begin{tabular}{l|ccp{9cm}} 
          \hline
            \textbf{Participants} & \textbf{Role} & \textbf{Source} & \textbf{Verbatim}
            \\ \hline
            \textbf{P2} & Influencer & During the game & I move only two circles from the right and from the left of the car. Only them, and the position and the size of them. Two dots which can be bigger or smaller. And it's like magnets, the cart is like magnet between them. \\ \hline 
            \textbf{P4} & Influencer & During the game & It seems like when the dots are really small the car is really calm. \\ \hline 
            \textbf{P6} & Influencer & Joint interview & The car was going faster when the car was not between the dots. \\ \hline 
            \textbf{P8} & Influencer & During the game & I think I understand something I think when you put force it pushes the vehicle to the side. \\ \hline 
            \textbf{P14} & Influencer & During the game & Maybe I can block the bar [pole] if I place the circles near the pole and not near the car. \\ \hline
            \textbf{P14} & Influencer & During the game & Maybe it’s about the difference of the distance between the circles and the car. Because I had an impression that when the car is in the middle of the circles it doesn’t move a lot. \\ \hline 
            \textbf{P24} & Influencer & During the game & The point of this game is to keep the car in the middle, as long as possible. \\ \hline 
            \textbf{P26} & Influencer & During the game & probably when I move the two circles, I think I control the speed of the car. \\ \hline
            \textbf{P32} & Influencer & Individual interview &My first impression was that the ball decided when to lose the game like if the car touches the ball it loses but not good intuition. \\

           \hline
        \end{tabular}
        
     \caption{Quotes of participants illustrating intuitions and interpretations participants had regarding the influences}
    \label{tab:quoteInterpretation}

\end{table}

\begin{table}
    \centering
        \begin{tabular}{l|ccp{9cm}} 
          \hline
            \textbf{Participants} & \textbf{Role} & \textbf{Source} & \textbf{Verbatim}
            \\ \hline
            \textbf{P4} & Influencer & Individual interview & When I did something really big, the cart went crazy very fast and left. \\ \hline 
            \textbf{P8} & Influencer & During the game & I have to manage, to control the force I put on right and left, in order to stabilise it. \\ \hline 
            \textbf{P10} & Influencer & During the game & I guess the circles are kind of \textit{attractor}. \\ \hline
             \textbf{P18} & Influencer & During the game & It works a little bit like a magnet you know \\ \hline \textbf{P24} & Influencer & Individual interview & It’s not really really like gravity. It’s like the opposite effect to the gravity. \\ \hline
            \textbf{P28} & Influencer & During the game & For me it’s the opposite of gravity effect, I increase the circles to move the car away. \\ \hline

            \textbf{P31} & Coach & During the game & maybe I don't know it's a hypothesis is but maybe the balls it's like a magnet you know ? \\  
            \textbf{P32} & Influencer & During the game & Yeah there like magnets I think, [...] like repealing magnets so when I will I make one ball bigger it pushes the car the other way. \\ \hline
            
            \textbf{P35} & Coach & During the game & Oh, so do you have some control of wind power ? \\  
            \textbf{P36} & Influencer & During the game & I think it's kind of a wind yes. \\ \hline
            
            \textbf{P44} & Influencer & During the game & yeah, I guess, I have some kind of control like with the forces that maybe attract or repulse the cart. \\ \hline
            
           \hline
        \end{tabular}
        
     \caption{Quotes of participants illustrating how influencers understood the impact of influences over the cartpole}
    \label{tab:quoteRepresentations}

\end{table}

In this last sequence of quotes in Table \ref{tab:quoteRepresentations}, we see that participants could elaborate a good understanding of the use of influences. 20 out of the 22 influencers involved in the experiment reported to feel having an effect on the cartpole with the influences.  Some of them could explain further how it modified the cartpole's behavior. We also collected from some influencers the feeling that they modified the cartpole's behaviour with the influences while not being able to make sense of how to control it. We note diverse ways to describe the role of influences (magnet, attractor, gravity, wind) and the elements of the game (gates, shapes, lines, objects, drops, tales, triangles). This illustrates that in the absence of information provided in introduction to the teams and each team elaborated their own vocabulary to discuss during the game.

The ways participants explained the effect of the influences are diverse and more or less precise. When explained as gravity or magnets it was very close to how we designed the influences. At times their explanations were not fully aligned with the influence's effects on the cartpole. What mattered here is that participants could find operating representations rather than correct representations. For instance, the two following quotes help in elaborating how to win rather than how influences work: "When I did something really big, the cart went crazy very fast and left" (P4) or "if the two circles have the good size, it's going itself toward the triangles [gates]" (P10).

In sum, we detail the learning curves of the teams in this section, our results suggest that their learning builds with the experience of the influences and results in an operating representation of how influences affect cartpole movements not necessarily requiring an accurate representation. This representation is in certain cases close to how it has been designed and would be described by the designers. In other cases, it is not correct. However, most influencers (20 out of 22) succeeded in developing a representation that allowed them to use the influences to achieve specific goals.


\subsection{Strategies to modify the position of the cartpole}
\label{sec:Strategies}

\begin{figure}
  \centering
  \includegraphics[width=1\columnwidth]{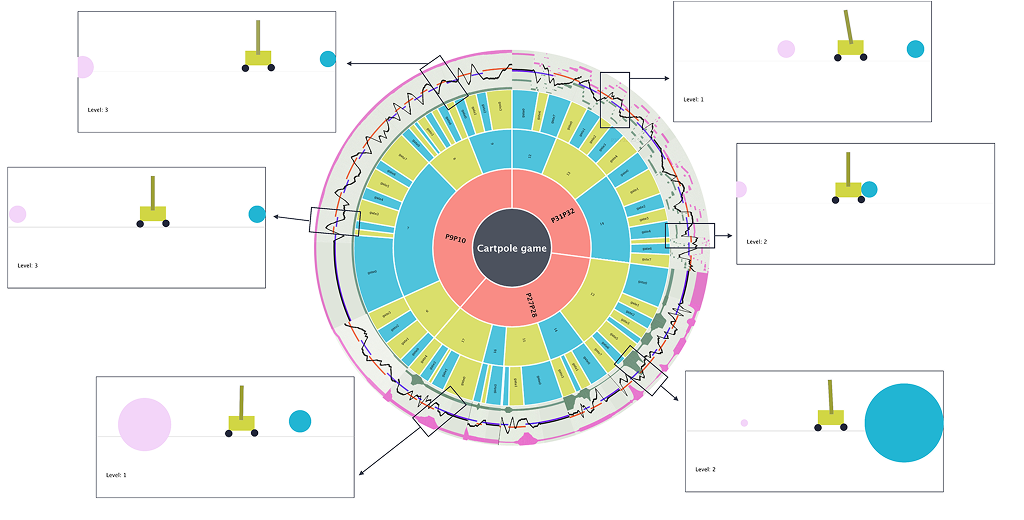 }
  \caption{This figure shows sections of the experiment of P31P32 (on the top right), P27P28 (on the bottom right), and P9P10 (on the left). It illustrates three different strategies to modify the cartpole position using the influences.}~\label{fig:Strategies}
\end{figure}

Among the high-level teams, we observed various ways of using the influences. In figure \ref{fig:Strategies} we illustrate the games of teams P9P10, P27P28, and P31P32, showing how they managed to control the position of the cartpole using the influences. Team P27P28 reached the highest number of passed gates in the experiment, passing 103 gates. The influencer P28 used influences by only modifying their size and not their position. P28 would balance the size of influences one after the other by making one larger and the other smaller. The cartpole would naturally move towards the smallest influence. This strategy proved to be very efficient in initiating the movement of the cartpole. This is a reason why P27P28 reached such a high level of the total number of passed gates (103). P27P28 also faced cartpole exits in several situations. In fact, when the cartpole went beyond the small influence, there was not enough time for the influencer to catch up to it.

  \vspace{-5pt}
The team P31P32 flanked the cartpole with the two influences and were moving both influences together to sort of "escort" the cartpole to the correct part of the screen. This strategy was effective, and required a lot of attention and reaction from the influencer even though P32 (the influencer) felt the cartpole to move slowly.

\begin{quote}
    P32 - Influencer \textit{during the game}: "it's not moving as easily [...] [to the cartpole]  Come on,  [to P31] he doesn’t want to go to the left. "
\end{quote}

P32 paid attention to never letting the cartpole exit from the two influences. This active control of the cartpole is visible in the percentage of influence in the cartpole actions (15.7 \%) which is much higher compared to P27P28 or P9P10 (respectively 5.99 \% and 4.81 \%) (See Table \ref{tab:TablePart1}). Here the influencer tried to actively control the cartpole. The necessary reactivity of P32 deployed given their strategy, helped them to often avoid exits of the screen or falls. 

Finally, P9P10 relied on much more static influences with little modifications. P10 relied on the natural "pendulum movement" of the cartpole in to limit the modifications of influences to the minimum.. 

\begin{quote}
    P10 - Influencer \textit{during the game}: "if the two circles have a good size, it's going itself toward the triangles [gates]."
\end{quote}

The position of the influences was efficiently preventing falls and exits, allowing the team to reach the highest score for average of \textit{ConsGates}. Without knowing it accurately, they allowed the model to do its job, while at the same time guiding it to not make the cartpole exit the screen, thus reached the best performance in our experiment.

From those three teams, we see that different use of influences can address different needs. For instance limiting the risks in avoiding exits and falls like (P9P10), increasing the speed when initiating the movement of the cartpole toward the desired goal while limiting the number of falls (P27P28), or reactively correcting the limits of the model in dynamically adapting the position and size of influences to avoid exits and falls (P31P32).

Given this game's context, our results demonstrate that 14 out of 22 teams reached a level of performance in less than 20 minutes without any prior information on the cartpole task or the functioning of influences. Our findings demonstrate how this operational understanding unfolds within different participant-created strategies, which proves to be efficient in achieving the goals of the game.


\section{Implications for Design}
\label{sec:ImplDesign}
\begin{figure}
  \centering
  \includegraphics[width=\textwidth]{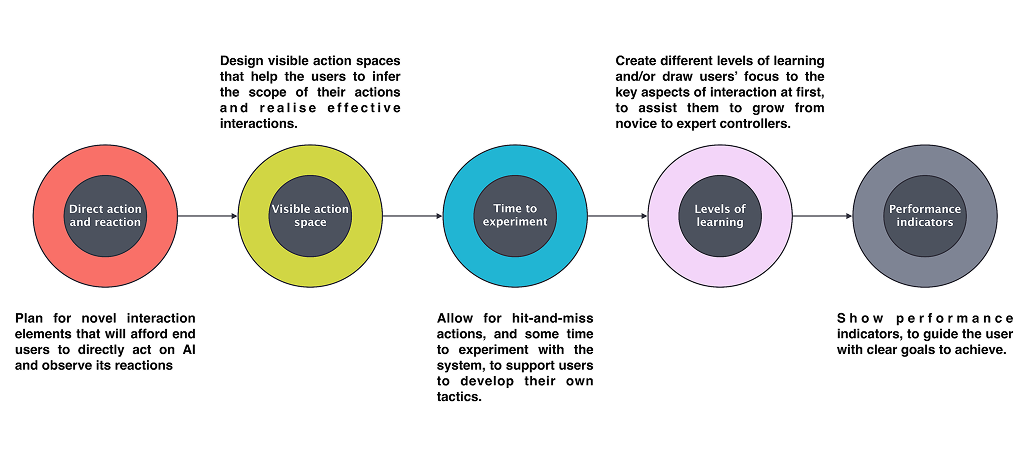}
  \caption{The five implications for designing actionable AI systems with (i) affordance for direct manipulation, (ii) visible action space defining the scope of user actions on the agent's abilities, (iii) time to experiment and fail before achieving success with operationalised understanding, (iv) different levels of learning with increasing levels of difficulties, and (v) clear performance indicators to drive users to succeed.}~\label{fig:implications}

\end{figure}

Based on our experimentation, featuring non-experts interacting with a black box agent, and our observations of emergent understanding leading to successful configuration in real-time, in this section, we derive implications for designing a framework for \textit{Actionable AI}  (Figure \ref{fig:implications}).




\subsection{Direct action and reaction} 
The real-time observation of the effects of direct manipulation is key to producing efficient learning in a limited amount of time. It is through this instant feedback loop that users can experiment with different options and actively develop their skills. In our work, the \textit{influences} allowed participants to have a direct impact on the ML model, which they immediately observed on the cartpole's movements. While we used influences because it suited our experimentation, any suitable interaction mechanism in any mode of interaction (such as virtual or embedded), that affords direct action on AI and leads to an observable reaction would serve as the primary step for making AI actionable. \textbf{Plan for novel interaction elements that will afford end users to directly act on AI and observe its reactions}.

\subsection{Visible action space} 
To effectively modify the behavior of an AI agent, beyond the observable effects of their action on AI, it is necessary to provide users with visible access to an actionable environment where they can clearly see the limits of the actions. Our participants could directly see the cartpole's playground. This perception gave them a quick account of the boundaries beyond which the cartpole crashes, so they saw that they had to be more careful in these regions. The single-dimensional nature of our game design made it easier for players to understand the agent's behavior and identify areas for improvement. However, in more complex environments with multiple dimensions, it can be more challenging to make the agent's action space accessible to users. For example, in a medical diagnostic setting, doctors would need to be able to see and understand the actions of a diagnostic AI model to effectively improve its performance. This raises the question of how to design actionable environments for agents operating in multi-dimensional spaces. While it may be possible to create physical environments for agents such as task-based robots and cars, it is more difficult to design actionable environments for non-physical tasks such as medical diagnosis and generative AI (such as large language models). In these cases, it may be necessary to create virtual environments that allow users to interact with the AI model and understand its action space in effectively configure its behavior. \textbf{Design visible action spaces that help the users to infer the scope of their actions and realise effective interactions}.


\subsection{Time to experiment} 
For \textit{Actionable AI}, the understanding of how to configure the agent arises from the time taken to try different actions. This entails added support with an experimentation phase, even for a short amount of time. In our experiments, the affordance we had created to play, by placing the \textit{influences} in different positions with respect to the cartpole, helped participants to gain an understanding of how it worked. Also, we chose to not share any instructions or clues, because we wanted to validate the extent to which it would be manageable for our participants to succeed in the limited time of 30 minutes. This safe space for experimentation can be for different amounts of time and have different levels of information shared with the user. \textbf{Allow for hit-and-miss actions, and some time to experiment with the system, to support users to develop their own tactics}.

 \subsection{Levels of learning} 
To increase the efficiency of the learning, the end-users can be guided through progressive levels of complexity. This is similar to curriculum learning. This does not necessarily imply to be time-consuming, but as designers of actionable solutions for AI, we might prefer that users first find out the major points of interaction before happening upon the subtleties for fine-tuning the agent. When participants started to succeed in our experiment, we promoted them to the next levels in the game, with an increase in difficulty with every level, which allowed them to refine their use of \textit{influences} and their understanding of the system. \textbf{Create different levels of learning with ascending levels of difficulty and/or draw users' focus to the key aspects of interaction at first, to assist them to grow from novice to expert in controlling the system}.

 \subsection{Performance indicators} 
Actionable AI aims at providing reliable outcomes such as a successful configuration of the agent, by indicating the performance of the agent-user collaboration. In our experiment, displaying the score of the game, gave participants concrete feedback on how they performed. Participants were goal-oriented and needed to find ways to improve their scores. It motivated them to maintain their engagement in the task. Performance indication can be a quantification of different elements, such as time to achieve the task or the quality of results obtained with the task. Without performance indicators, it would be difficult for the users to discern from clear intents or their control over the system. \textbf{Show performance indicators, to guide the user with clear goals to achieve}.

  \vspace{-5pt}

\section{Discussion}

\subsection{Human-AI Collaboration}
As described in section \ref{sec:Strategies} teams relied on different ways to use influences. From the high-level teams in the experiments P9P10 and P27P28, the cartpole followed the influence between 5\% and 6\% of the steps. For P3132, also a high-level team, the cartpole followed the influences around 15\% of the steps. These differences in influence usage for developing successful strategies illustrate different practices of the influencer. For the case of P9P10 and P27P28, we can interpret that the teams considered the cartpole behavior as an asset to exploit and included it in their monitoring. This was relevant as the task of the model controlling the cartpole was to maintain the pole on the cart and that was a target teams also had to maintain to avoid the pole from falling. These teams exhibited a strong understanding of the cartpole's behaviour by going along with the model for succeeding in their goal. On the other hand, P32 really tried to impose a new behaviour on the cartpole using the influences, where P32 was constantly pushing and pulling the cartpole, feeling that the cartpole exhibited a slow-moving behaviour. Certainly, there is a limit to the temptation to override the behaviour of the model with the influences and it even comes with some risks to break the model and decrease its performance. The strategies employed by P9P10 or P27P28 appear to be more efficient where they were slightly adjusting the movements of the cartpole only when necessary, for instance at the beginning of a new gate to initiate the movement of the cartpole in the right direction. Our findings add to the evidence base for building a possible future where humans seamlessly collaborate with AI systems as envisioned by the pioneers of Human-Centred AI \cite{haicollab}.

\subsection{Levels of explanations}
\label{sec:DiscExplanation}
We purposefully designed an experiment where participants had no clue about the action of the cartpole (the goal of the ML model) and the effect of the influences on the cartpole. We designed it this way to assess the possibility of using influences for non-experts and non-informed participants. We also wanted to observe if and how the understanding of the cartpole, with influences would develop during the experiment. We could have chosen to give more explanations to the end users. First, by describing the goal of the model controlling the cartpole which was to adjust the movements of the cart to prevent the pole from falling. Then we could have described that influences acted like magnets on the cartpole, i.e. they applied a directional force quite similar to the opposite effect of gravity on the cartpole, that would push the cartpole away depending on their size and distance to the cart. This description is close to what we reported in our findings (Section \ref{sec:findings}) as the one shared by many participants. Nevertheless, this description is not accurate. What the influences do indeed is attract the cart, exactly the very same as a gravitational effect. The reason why the cartpole seems to move in the opposite direction, (thus exhibiting anti-gravity) was because this initial movement of the cart led to unbalancing of the pole, and the model which directly corrected this unbalancing tried to balance the pole again by moving the cart in the opposite direction. This is the true level of explanation of how the influences worked with the cartpole, this is misleading and complex. In practice, if shared with users, it might confuse them on how to use the influences. Realistically, the most valuable explanation that helps to configure the AI is what we identified as the operating representation: An explanation that is usable in practice. Additionally, the randomness or uncertainty of the black-box model was unearthed by our participants as evidently seen in Table \ref{tab:quoteDoubts}, where they express their doubts about the random functioning of the cartpole. This adheres to the ethical guidelines for building AI systems \cite{ethicalGuidelines}, rather than explaining the unexplainable randomness of black-box models. We do not say that Actionable AI is a replace Explainable AI, nor do we recommend that users should not be given explanations, we envision Actionable AI to be used in parallel, say for gaining an operational knowledge of the ML model, with Explainable AI, say for communicating the intended use of the system and data usage. 

Given the science-fiction-fuelled current mental models of everyday users on AI \cite{DBLP:journals/corr/abs-2011-02731}, we studied whether the information \textit{an AI algorithm is controlling the cartpole} had any impact on performances, or ways to make sense of the cartpole game during the game. For instance, the knowledgeable teams could have been possibly more confused because of the currents state of misleading AI promises \cite{misAI}, or on the contrary, more accurate in an understanding of the cartpole aim. None of this had an impact on our findings which suggest that through Actionable AI what takes over is the active position the user has as most of the participants (21 out of 22) who were informed of the AI, reported during their joint interview to forget the involvement of AI, as they were actively involved in playing the game.
This resulted in actively interacting with the influences, without sustaining preconceived biases about the AI, to make sense of the joint system \textit{influences and cartpole} through reflective practices \cite{mollo2014reflective}.

\section{Limitations and Future Work}
The cartpole game used in this contribution consist of a simple task with decisions limited to a single dimension. Our contribution is a preliminary experiment that allowed us to answer our research question. Further research is required to study different agents (beyond navigation), that can be configurable with \textit{influences} or any other direct manipulation interaction mechanism.

We modified Open AI's single-user cartpole game \cite{OpenAIGym} into a collaborative setup, focusing on two-player interactions. This narrows the scope to specific conversational dynamics, limiting broader applicability to other AI interaction scenarios, such as one-on-one or multi-user engagements. Further research is needed to explore these diverse interaction models.

To explore the potential of Actionable AI framework, further experiments should address the impact of a more complex environment such as navigating objects in real environment. Further work is needed to explore optimisation of the time needed by everyday users to tune black-blox AI, and the effect of actionable AI framework on user experience and adoption of AI.



\section{Conclusion}
Human-AI interaction can be viewed as AI serving the human by automating a task, or as humans serving the AI with the Human-in-the-Loop improving and training of ML algorithms. With Actionable AI we propose another path, to leverage direct manipulation tools from the Human-in-the-Loop research field, by making them available to non-experts as interaction instruments, to enable the configuration of AI systems via operational understanding in uncertain conditions. In this paper, we demonstrate Actionable AI with a cartpole gaming experiment, with 22 pairs of participants. Our findings show that through our framework, 14 out of 22 teams succeeded to configure the AI system toward their goals, and 20 out of 22 players who controlled the system gained an understanding of it, which was efficient in achieving their goals and employing different strategies for the same. Based on our experiment, we derive implications for Actionable AI, which is constructed with (1) affording direct manipulation of the AI agent, (2) providing the user with a visible action space, (3) giving users the time to experiment, (4)  designing for the users to gain different levels of operational understanding, and (5) resulting in performance indicators of a successful configuration of the AI system.

\begin{acks}
We thank our participants for their time and part; and our colleagues for their feedback on earlier versions of this paper.
\end{acks}

\bibliographystyle{ACM-Reference-Format}
\bibliography{bib}


\begin{thebibliography}{48}


\ifx \showCODEN    \undefined \def \showCODEN     #1{\unskip}     \fi
\ifx \showDOI      \undefined \def \showDOI       #1{#1}\fi
\ifx \showISBNx    \undefined \def \showISBNx     #1{\unskip}     \fi
\ifx \showISBNxiii \undefined \def \showISBNxiii  #1{\unskip}     \fi
\ifx \showISSN     \undefined \def \showISSN      #1{\unskip}     \fi
\ifx \showLCCN     \undefined \def \showLCCN      #1{\unskip}     \fi
\ifx \shownote     \undefined \def \shownote      #1{#1}          \fi
\ifx \showarticletitle \undefined \def \showarticletitle #1{#1}   \fi
\ifx \showURL      \undefined \def \showURL       {\relax}        \fi
\providecommand\bibfield[2]{#2}
\providecommand\bibinfo[2]{#2}
\providecommand\natexlab[1]{#1}
\providecommand\showeprint[2][]{arXiv:#2}

\bibitem[Ackerman(2020)]%
        {Ackerman}
\bibfield{author}{\bibinfo{person}{Evan Ackerman}.} \bibinfo{year}{25 Aug 2020}\natexlab{}.
\newblock \bibinfo{title}{iRobot Announces Major Software Update, Shift From Pure Autonomy to Human-Robot Collaboration}.
\newblock \bibinfo{howpublished}{\url{https://spectrum.ieee.org/irobot-home-autonomy-update}}. , \bibinfo{numpages}{117--133}~pages.
\newblock


\bibitem[AI(2006)]%
        {OpenAIGym}
\bibfield{author}{\bibinfo{person}{Open AI}.} \bibinfo{year}{2006}\natexlab{}.
\newblock \bibinfo{title}{Cartpole OpenAI Gym}.
\newblock
\newblock
\urldef\tempurl%
\url{https://github.com/openai/gym/blob/master/gym/envs/classic_control/cartpole.py}
\showURL{%
Retrieved August, 25, 2022 from \tempurl}


\bibitem[AI(2021)]%
        {Dalle2}
\bibfield{author}{\bibinfo{person}{Open AI}.} \bibinfo{year}{2021}\natexlab{}.
\newblock \bibinfo{title}{DALL·E}.
\newblock
\newblock
\urldef\tempurl%
\url{https://openai.com/dall-e-2/}
\showURL{%
Retrieved December, 10, 2022 from \tempurl}


\bibitem[AI(2022)]%
        {chatgpt}
\bibfield{author}{\bibinfo{person}{Open AI}.} \bibinfo{year}{2022}\natexlab{}.
\newblock \bibinfo{title}{ChatGPT: Optimizing Language Models for Dialogue}.
\newblock
\newblock
\urldef\tempurl%
\url{https://openai.com/blog/chatgpt/}
\showURL{%
Retrieved December, 10, 2022 from \tempurl}


\bibitem[Arrieta et~al\mbox{.}(2020)]%
        {arrieta2020explainable}
\bibfield{author}{\bibinfo{person}{Alejandro~Barredo Arrieta}, \bibinfo{person}{Natalia D{\'\i}az-Rodr{\'\i}guez}, \bibinfo{person}{Javier Del~Ser}, \bibinfo{person}{Adrien Bennetot}, \bibinfo{person}{Siham Tabik}, \bibinfo{person}{Alberto Barbado}, \bibinfo{person}{Salvador Garc{\'\i}a}, \bibinfo{person}{Sergio Gil-L{\'o}pez}, \bibinfo{person}{Daniel Molina}, \bibinfo{person}{Richard Benjamins}, {et~al\mbox{.}}} \bibinfo{year}{2020}\natexlab{}.
\newblock \showarticletitle{Explainable Artificial Intelligence (XAI): Concepts, taxonomies, opportunities and challenges toward responsible AI}.
\newblock \bibinfo{journal}{\emph{Information fusion}}  \bibinfo{volume}{58} (\bibinfo{year}{2020}), \bibinfo{pages}{82--115}.
\newblock


\bibitem[Ashktorab et~al\mbox{.}(2021)]%
        {directionality}
\bibfield{author}{\bibinfo{person}{Zahra Ashktorab}, \bibinfo{person}{Casey Dugan}, \bibinfo{person}{James Johnson}, \bibinfo{person}{Qian Pan}, \bibinfo{person}{Wei Zhang}, \bibinfo{person}{Sadhana Kumaravel}, {and} \bibinfo{person}{Murray Campbell}.} \bibinfo{year}{2021}\natexlab{}.
\newblock \showarticletitle{Effects of Communication Directionality and AI Agent Differences in Human-AI Interaction}. In \bibinfo{booktitle}{\emph{Proceedings of the 2021 CHI Conference on Human Factors in Computing Systems}} (Yokohama, Japan) \emph{(\bibinfo{series}{CHI '21})}. \bibinfo{publisher}{Association for Computing Machinery}, \bibinfo{address}{New York, NY, USA}, Article \bibinfo{articleno}{238}, \bibinfo{numpages}{15}~pages.
\newblock
\showISBNx{9781450380966}
\urldef\tempurl%
\url{https://doi.org/10.1145/3411764.3445256}
\showDOI{\tempurl}


\bibitem[Atlantic(2022)]%
        {atlantic}
\bibfield{author}{\bibinfo{person}{The Atlantic}.} \bibinfo{year}{2022}\natexlab{}.
\newblock \bibinfo{title}{hatGPT gained 1 million users in under a week. Here’s why the AI chatbot is primed to disrupt search as we know it}.
\newblock
\newblock
\urldef\tempurl%
\url{https://www.theatlantic.com/technology/archive/2022/12/openai-chatgpt-chatbot-messages/672411/}
\showURL{%
Retrieved December, 10, 2022 from \tempurl}


\bibitem[Balawejder(2021)]%
        {Cartpole}
\bibfield{author}{\bibinfo{person}{Maciej Balawejder}.} \bibinfo{year}{2021}\natexlab{}.
\newblock \showarticletitle{Solving Open AI’s CartPole Using Reinforcement Learning}.
\newblock  (\bibinfo{year}{2021}).
\newblock
\urldef\tempurl%
\url{https://medium.com/analytics-vidhya/q-learning-is-the-most-basic-form-of-reinforcement-learning-which-doesnt-take-advantage-of-any-8944e02570c5}
\showURL{%
Retrieved August, 25, 2022 from \tempurl}


\bibitem[Braun and Clarke(2006)]%
        {thematicAnalysis}
\bibfield{author}{\bibinfo{person}{Virginia Braun} {and} \bibinfo{person}{Victoria Clarke}.} \bibinfo{year}{2006}\natexlab{}.
\newblock \showarticletitle{Using thematic analysis in psychology}.
\newblock \bibinfo{journal}{\emph{Qualitative Research in Psychology}} \bibinfo{volume}{3}, \bibinfo{number}{2} (\bibinfo{year}{2006}), \bibinfo{pages}{77--101}.
\newblock
\urldef\tempurl%
\url{https://doi.org/10.1191/1478088706qp063oa}
\showDOI{\tempurl}
\showeprint{https://www.tandfonline.com/doi/pdf/10.1191/1478088706qp063oa}


\bibitem[Brennen(2020)]%
        {10.1145/3334480.3383047}
\bibfield{author}{\bibinfo{person}{Andrea Brennen}.} \bibinfo{year}{2020}\natexlab{}.
\newblock \showarticletitle{What Do People Really Want When They Say They Want "Explainable AI?" We Asked 60 Stakeholders.}. In \bibinfo{booktitle}{\emph{Extended Abstracts of the 2020 CHI Conference on Human Factors in Computing Systems}} (Honolulu, HI, USA) \emph{(\bibinfo{series}{CHI EA '20})}. \bibinfo{publisher}{Association for Computing Machinery}, \bibinfo{address}{New York, NY, USA}, \bibinfo{pages}{1–7}.
\newblock
\showISBNx{9781450368193}
\urldef\tempurl%
\url{https://doi.org/10.1145/3334480.3383047}
\showDOI{\tempurl}


\bibitem[Buro(2003)]%
        {RTS}
\bibfield{author}{\bibinfo{person}{Michael Buro}.} \bibinfo{year}{2003}\natexlab{}.
\newblock \showarticletitle{Real-Time Strategy Games: A New AI Research Challenge.} \bibinfo{pages}{1534--1535}.
\newblock


\bibitem[Cakmak et~al\mbox{.}(2010)]%
        {cakmak2010designing}
\bibfield{author}{\bibinfo{person}{Maya Cakmak}, \bibinfo{person}{Crystal Chao}, {and} \bibinfo{person}{Andrea~L Thomaz}.} \bibinfo{year}{2010}\natexlab{}.
\newblock \showarticletitle{Designing interactions for robot active learners}.
\newblock \bibinfo{journal}{\emph{IEEE Transactions on Autonomous Mental Development}} \bibinfo{volume}{2}, \bibinfo{number}{2} (\bibinfo{year}{2010}), \bibinfo{pages}{108--118}.
\newblock


\bibitem[Caldwell et~al\mbox{.}(2022)]%
        {transferability}
\bibfield{author}{\bibinfo{person}{Sabrina Caldwell}, \bibinfo{person}{Penny Sweetser}, \bibinfo{person}{Nicholas O’Donnell}, \bibinfo{person}{Matthew~J. Knight}, \bibinfo{person}{Matthew Aitchison}, \bibinfo{person}{Tom Gedeon}, \bibinfo{person}{Daniel Johnson}, \bibinfo{person}{Margot Brereton}, \bibinfo{person}{Marcus Gallagher}, {and} \bibinfo{person}{David Conroy}.} \bibinfo{year}{2022}\natexlab{}.
\newblock \showarticletitle{An Agile New Research Framework for Hybrid Human-AI Teaming: Trust, Transparency, and Transferability}.
\newblock \bibinfo{journal}{\emph{ACM Trans. Interact. Intell. Syst.}} \bibinfo{volume}{12}, \bibinfo{number}{3}, Article \bibinfo{articleno}{17} (\bibinfo{date}{jul} \bibinfo{year}{2022}), \bibinfo{numpages}{36}~pages.
\newblock
\showISSN{2160-6455}
\urldef\tempurl%
\url{https://doi.org/10.1145/3514257}
\showDOI{\tempurl}


\bibitem[Chao et~al\mbox{.}(2010)]%
        {chao2010transparent}
\bibfield{author}{\bibinfo{person}{Crystal Chao}, \bibinfo{person}{Maya Cakmak}, {and} \bibinfo{person}{Andrea~L Thomaz}.} \bibinfo{year}{2010}\natexlab{}.
\newblock \showarticletitle{Transparent active learning for robots}. In \bibinfo{booktitle}{\emph{2010 5th ACM/IEEE International Conference on Human-Robot Interaction (HRI)}}. IEEE, \bibinfo{pages}{317--324}.
\newblock


\bibitem[D3.js(2011)]%
        {D3}
\bibfield{author}{\bibinfo{person}{D3.js}.} \bibinfo{year}{2011}\natexlab{}.
\newblock \bibinfo{title}{Data-Driven Documents}.
\newblock
\newblock
\urldef\tempurl%
\url{https://d3js.org}
\showURL{%
\tempurl}


\bibitem[Dang et~al\mbox{.}(2022)]%
        {dang2022prompt}
\bibfield{author}{\bibinfo{person}{Hai Dang}, \bibinfo{person}{Lukas Mecke}, \bibinfo{person}{Florian Lehmann}, \bibinfo{person}{Sven Goller}, {and} \bibinfo{person}{Daniel Buschek}.} \bibinfo{year}{2022}\natexlab{}.
\newblock \showarticletitle{How to Prompt? Opportunities and Challenges of Zero-and Few-Shot Learning for Human-AI Interaction in Creative Applications of Generative Models}.
\newblock \bibinfo{journal}{\emph{arXiv preprint arXiv:2209.01390}} (\bibinfo{year}{2022}).
\newblock


\bibitem[Das et~al\mbox{.}(2021)]%
        {RobotExplanation}
\bibfield{author}{\bibinfo{person}{Devleena Das}, \bibinfo{person}{Siddhartha Banerjee}, {and} \bibinfo{person}{Sonia Chernova}.} \bibinfo{year}{2021}\natexlab{}.
\newblock \showarticletitle{Explainable AI for Robot Failures: Generating Explanations That Improve User Assistance in Fault Recovery}. In \bibinfo{booktitle}{\emph{Proceedings of the 2021 ACM/IEEE International Conference on Human-Robot Interaction}} (Boulder, CO, USA) \emph{(\bibinfo{series}{HRI '21})}. \bibinfo{publisher}{Association for Computing Machinery}, \bibinfo{address}{New York, NY, USA}, \bibinfo{pages}{351–360}.
\newblock
\showISBNx{9781450382892}
\urldef\tempurl%
\url{https://doi.org/10.1145/3434073.3444657}
\showDOI{\tempurl}


\bibitem[Doran et~al\mbox{.}(2017)]%
        {https://doi.org/10.48550/arxiv.1710.00794}
\bibfield{author}{\bibinfo{person}{Derek Doran}, \bibinfo{person}{Sarah Schulz}, {and} \bibinfo{person}{Tarek~R. Besold}.} \bibinfo{year}{2017}\natexlab{}.
\newblock \bibinfo{title}{What Does Explainable AI Really Mean? A New Conceptualization of Perspectives}.
\newblock
\newblock
\urldef\tempurl%
\url{https://doi.org/10.48550/ARXIV.1710.00794}
\showDOI{\tempurl}


\bibitem[Dourish(2003)]%
        {dourish2003adoptadapt}
\bibfield{author}{\bibinfo{person}{Paul Dourish}.} \bibinfo{year}{2003}\natexlab{}.
\newblock \showarticletitle{The appropriation of interactive technologies: Some lessons from placeless documents}.
\newblock \bibinfo{journal}{\emph{Computer Supported Cooperative Work (CSCW)}} \bibinfo{volume}{12}, \bibinfo{number}{4} (\bibinfo{year}{2003}), \bibinfo{pages}{465--490}.
\newblock


\bibitem[Dudley and Kristensson(2018)]%
        {IML}
\bibfield{author}{\bibinfo{person}{John~J. Dudley} {and} \bibinfo{person}{Per~Ola Kristensson}.} \bibinfo{year}{2018}\natexlab{}.
\newblock \showarticletitle{A Review of User Interface Design for Interactive Machine Learning}.
\newblock \bibinfo{journal}{\emph{ACM Trans. Interact. Intell. Syst.}} \bibinfo{volume}{8}, \bibinfo{number}{2}, Article \bibinfo{articleno}{8} (\bibinfo{date}{jun} \bibinfo{year}{2018}), \bibinfo{numpages}{37}~pages.
\newblock
\showISSN{2160-6455}
\urldef\tempurl%
\url{https://doi.org/10.1145/3185517}
\showDOI{\tempurl}


\bibitem[Fiebrink et~al\mbox{.}(2009)]%
        {fiebrink2009meta}
\bibfield{author}{\bibinfo{person}{Rebecca Fiebrink}, \bibinfo{person}{Daniel Trueman}, \bibinfo{person}{Perry~R Cook}, {et~al\mbox{.}}} \bibinfo{year}{2009}\natexlab{}.
\newblock \showarticletitle{A meta-instrument for interactive, on-the-fly machine learning}.
\newblock  (\bibinfo{year}{2009}).
\newblock


\bibitem[Finance(2022)]%
        {chatgptUsers}
\bibfield{author}{\bibinfo{person}{Yahoon Finance}.} \bibinfo{year}{2022}\natexlab{}.
\newblock \bibinfo{title}{Five Remarkable Chats That Will Help You Understand ChatGPT}.
\newblock
\newblock
\urldef\tempurl%
\url{https://finance.yahoo.com/news/chatgpt-gained-1-million-followers-224523258.html}
\showURL{%
Retrieved December, 10, 2022 from \tempurl}


\bibitem[Gao et~al\mbox{.}(2021)]%
        {Gao2021HumanAICW}
\bibfield{author}{\bibinfo{person}{Ruijiang Gao}, \bibinfo{person}{Maytal Saar-Tsechansky}, \bibinfo{person}{Maria De-Arteaga}, \bibinfo{person}{Ligong Han}, \bibinfo{person}{Min~Kyung Lee}, {and} \bibinfo{person}{Matthew Lease}.} \bibinfo{year}{2021}\natexlab{}.
\newblock \showarticletitle{Human-AI Collaboration with Bandit Feedback}. In \bibinfo{booktitle}{\emph{International Joint Conference on Artificial Intelligence}}.
\newblock


\bibitem[Gillies et~al\mbox{.}(2015)]%
        {IMLDebugging}
\bibfield{author}{\bibinfo{person}{Marco Gillies}, \bibinfo{person}{Andrea Kleinsmith}, {and} \bibinfo{person}{Harry Brenton}.} \bibinfo{year}{2015}\natexlab{}.
\newblock \showarticletitle{Applying the CASSM Framework to Improving End User Debugging of Interactive Machine Learning}. In \bibinfo{booktitle}{\emph{Proceedings of the 20th International Conference on Intelligent User Interfaces}} (Atlanta, Georgia, USA) \emph{(\bibinfo{series}{IUI '15})}. \bibinfo{publisher}{Association for Computing Machinery}, \bibinfo{address}{New York, NY, USA}, \bibinfo{pages}{181–185}.
\newblock
\showISBNx{9781450333061}
\urldef\tempurl%
\url{https://doi.org/10.1145/2678025.2701373}
\showDOI{\tempurl}


\bibitem[G{\"o}nen(2008)]%
        {gonen2008study}
\bibfield{author}{\bibinfo{person}{Selahattin G{\"o}nen}.} \bibinfo{year}{2008}\natexlab{}.
\newblock \showarticletitle{A study on student teachers’ misconceptions and scientifically acceptable conceptions about mass and gravity}.
\newblock \bibinfo{journal}{\emph{Journal of Science Education and Technology}} \bibinfo{volume}{17}, \bibinfo{number}{1} (\bibinfo{year}{2008}), \bibinfo{pages}{70--81}.
\newblock


\bibitem[Gunning et~al\mbox{.}(2019)]%
        {gunning2019xai}
\bibfield{author}{\bibinfo{person}{David Gunning}, \bibinfo{person}{Mark Stefik}, \bibinfo{person}{Jaesik Choi}, \bibinfo{person}{Timothy Miller}, \bibinfo{person}{Simone Stumpf}, {and} \bibinfo{person}{Guang-Zhong Yang}.} \bibinfo{year}{2019}\natexlab{}.
\newblock \showarticletitle{XAI—Explainable artificial intelligence}.
\newblock \bibinfo{journal}{\emph{Science Robotics}} \bibinfo{volume}{4}, \bibinfo{number}{37} (\bibinfo{year}{2019}), \bibinfo{pages}{eaay7120}.
\newblock


\bibitem[Guzdial et~al\mbox{.}(2019)]%
        {collabRoles}
\bibfield{author}{\bibinfo{person}{Matthew Guzdial}, \bibinfo{person}{Nicholas Liao}, \bibinfo{person}{Jonathan Chen}, \bibinfo{person}{Shao-Yu Chen}, \bibinfo{person}{Shukan Shah}, \bibinfo{person}{Vishwa Shah}, \bibinfo{person}{Joshua Reno}, \bibinfo{person}{Gillian Smith}, {and} \bibinfo{person}{Mark~O. Riedl}.} \bibinfo{year}{2019}\natexlab{}.
\newblock \showarticletitle{Friend, Collaborator, Student, Manager: How Design of an AI-Driven Game Level Editor Affects Creators}. In \bibinfo{booktitle}{\emph{Proceedings of the 2019 CHI Conference on Human Factors in Computing Systems}} (Glasgow, Scotland Uk) \emph{(\bibinfo{series}{CHI '19})}. \bibinfo{publisher}{Association for Computing Machinery}, \bibinfo{address}{New York, NY, USA}, \bibinfo{pages}{1–13}.
\newblock
\showISBNx{9781450359702}
\urldef\tempurl%
\url{https://doi.org/10.1145/3290605.3300854}
\showDOI{\tempurl}


\bibitem[Hindemith et~al\mbox{.}(2020)]%
        {DBLP:journals/corr/abs-2011-02731}
\bibfield{author}{\bibinfo{person}{Lukas Hindemith}, \bibinfo{person}{Anna{-}Lisa Vollmer}, \bibinfo{person}{Jan~Philip G{\"{o}}pfert}, \bibinfo{person}{Christiane~B. Wiebel{-}Herboth}, {and} \bibinfo{person}{Britta Wrede}.} \bibinfo{year}{2020}\natexlab{}.
\newblock \showarticletitle{Why robots should be technical: Correcting mental models through technical architecture concepts}.
\newblock \bibinfo{journal}{\emph{CoRR}}  \bibinfo{volume}{abs/2011.02731} (\bibinfo{year}{2020}).
\newblock
\showeprint[arXiv]{2011.02731}
\urldef\tempurl%
\url{https://arxiv.org/abs/2011.02731}
\showURL{%
\tempurl}


\bibitem[Holzinger(2018)]%
        {8490530}
\bibfield{author}{\bibinfo{person}{Andreas Holzinger}.} \bibinfo{year}{2018}\natexlab{}.
\newblock \showarticletitle{From Machine Learning to Explainable AI}. In \bibinfo{booktitle}{\emph{2018 World Symposium on Digital Intelligence for Systems and Machines (DISA)}}. \bibinfo{pages}{55--66}.
\newblock
\urldef\tempurl%
\url{https://doi.org/10.1109/DISA.2018.8490530}
\showDOI{\tempurl}


\bibitem[Jacquin et~al\mbox{.}(2022)]%
        {JacquinPerezBoulard}
\bibfield{author}{\bibinfo{person}{Thierry Jacquin}, \bibinfo{person}{Julien Perez}, {and} \bibinfo{person}{Cecile Boulard}.} \bibinfo{year}{2022}\natexlab{}.
\newblock \showarticletitle{Human Influence in the Lifelong Reinforcement Learning Loop}. In \bibinfo{booktitle}{\emph{Proceedings of Lifelong Learning and Personalization in Long-Term Human-Robot Interaction (LEAP-HRI), Workshop at HRI 2022}}.
\newblock


\bibitem[Kokkinakis et~al\mbox{.}(2017)]%
        {fluidIntel}
\bibfield{author}{\bibinfo{person}{Athanasios~V. Kokkinakis}, \bibinfo{person}{Peter~I. Cowling}, \bibinfo{person}{Anders Drachen}, {and} \bibinfo{person}{Alex~R. Wade}.} \bibinfo{year}{2017}\natexlab{}.
\newblock \showarticletitle{Exploring the relationship between video game expertise and fluid intelligence}.
\newblock \bibinfo{journal}{\emph{PLOS ONE}} \bibinfo{volume}{12}, \bibinfo{number}{11} (\bibinfo{date}{11} \bibinfo{year}{2017}), \bibinfo{pages}{1--15}.
\newblock
\urldef\tempurl%
\url{https://doi.org/10.1371/journal.pone.0186621}
\showDOI{\tempurl}


\bibitem[Liang et~al\mbox{.}(2019)]%
        {liang2019implicit}
\bibfield{author}{\bibinfo{person}{Claire Liang}, \bibinfo{person}{Julia Proft}, \bibinfo{person}{Erik Andersen}, {and} \bibinfo{person}{Ross~A Knepper}.} \bibinfo{year}{2019}\natexlab{}.
\newblock \showarticletitle{Implicit communication of actionable information in human-ai teams}. In \bibinfo{booktitle}{\emph{Proceedings of the 2019 CHI Conference on Human Factors in Computing Systems}}. \bibinfo{pages}{1--13}.
\newblock


\bibitem[Miller(2023)]%
        {miller2023explainableaideadlong}
\bibfield{author}{\bibinfo{person}{Tim Miller}.} \bibinfo{year}{2023}\natexlab{}.
\newblock \bibinfo{title}{Explainable AI is Dead, Long Live Explainable AI! Hypothesis-driven decision support}.
\newblock
\newblock
\showeprint[arxiv]{2302.12389}~[cs.AI]
\urldef\tempurl%
\url{https://arxiv.org/abs/2302.12389}
\showURL{%
\tempurl}


\bibitem[Mollo and Nascimento(2014)]%
        {mollo2014reflective}
\bibfield{author}{\bibinfo{person}{Vanina Mollo} {and} \bibinfo{person}{Adelaide Nascimento}.} \bibinfo{year}{2014}\natexlab{}.
\newblock \showarticletitle{Reflective practices and the development of individuals, collectives and organizations}.
\newblock \bibinfo{journal}{\emph{Constructive ergonomics}}  \bibinfo{volume}{15} (\bibinfo{year}{2014}), \bibinfo{pages}{223--238}.
\newblock


\bibitem[Pinhanez(2021)]%
        {ethicalGuidelines}
\bibfield{author}{\bibinfo{person}{Claudio~S. Pinhanez}.} \bibinfo{year}{2021}\natexlab{}.
\newblock \bibinfo{title}{Expose Uncertainty, Instill Distrust, Avoid Explanations: Towards Ethical Guidelines for AI}.
\newblock
\newblock
\urldef\tempurl%
\url{https://doi.org/10.48550/ARXIV.2112.01281}
\showDOI{\tempurl}


\bibitem[Ramos et~al\mbox{.}(2020)]%
        {IMT}
\bibfield{author}{\bibinfo{person}{Gonzalo Ramos}, \bibinfo{person}{Christopher Meek}, \bibinfo{person}{Patrice Simard}, \bibinfo{person}{Jina Suh}, {and} \bibinfo{person}{Soroush Ghorashi}.} \bibinfo{year}{2020}\natexlab{}.
\newblock \showarticletitle{Interactive machine teaching: a human-centered approach to building machine-learned models}.
\newblock \bibinfo{journal}{\emph{Human‚ÄìComputer Interaction}} \bibinfo{volume}{35}, \bibinfo{number}{5-6} (\bibinfo{year}{2020}), \bibinfo{pages}{413--451}.
\newblock
\urldef\tempurl%
\url{https://doi.org/10.1080/07370024.2020.1734931}
\showDOI{\tempurl}
\showeprint{https://doi.org/10.1080/07370024.2020.1734931}


\bibitem[Rezwana and Maher(2022)]%
        {userperceptions}
\bibfield{author}{\bibinfo{person}{Jeba Rezwana} {and} \bibinfo{person}{Mary~Lou Maher}.} \bibinfo{year}{2022}\natexlab{}.
\newblock \showarticletitle{Understanding User Perceptions, Collaborative Experience and User Engagement in Different Human-AI Interaction Designs for Co-Creative Systems}. In \bibinfo{booktitle}{\emph{Creativity and Cognition}} (Venice, Italy). \bibinfo{publisher}{Association for Computing Machinery}, \bibinfo{address}{New York, NY, USA}, \bibinfo{pages}{38–48}.
\newblock
\showISBNx{9781450393270}
\urldef\tempurl%
\url{https://doi.org/10.1145/3527927.3532789}
\showDOI{\tempurl}


\bibitem[Settles(2009)]%
        {settles2009active}
\bibfield{author}{\bibinfo{person}{Burr Settles}.} \bibinfo{year}{2009}\natexlab{}.
\newblock \showarticletitle{Active learning literature survey}.
\newblock  (\bibinfo{year}{2009}).
\newblock


\bibitem[Shneiderman(2022)]%
        {shneiderman2022human}
\bibfield{author}{\bibinfo{person}{Ben Shneiderman}.} \bibinfo{year}{2022}\natexlab{}.
\newblock \bibinfo{booktitle}{\emph{Human-Centered AI}}.
\newblock \bibinfo{publisher}{Oxford University Press}.
\newblock


\bibitem[Tucker(2022)]%
        {misAI}
\bibfield{author}{\bibinfo{person}{Emily Tucker}.} \bibinfo{year}{2022}\natexlab{}.
\newblock \bibinfo{title}{AI}.
\newblock
\newblock
\urldef\tempurl%
\url{https://techpolicy.press/artifice-and-intelligence/}
\showURL{%
\tempurl}


\bibitem[Viswanathan et~al\mbox{.}(2022)]%
        {SitRec}
\bibfield{author}{\bibinfo{person}{Sruthi Viswanathan}, \bibinfo{person}{Cecile Boulard}, \bibinfo{person}{Adrien Bruyat}, {and} \bibinfo{person}{Antonietta Maria~Grasso}.} \bibinfo{year}{2022}\natexlab{}.
\newblock \showarticletitle{Situational Recommender: Are You On the Spot, Refining Plans, or Just Bored?}. In \bibinfo{booktitle}{\emph{Proceedings of the 2022 CHI Conference on Human Factors in Computing Systems}} (New Orleans, LA, USA) \emph{(\bibinfo{series}{CHI '22})}. \bibinfo{publisher}{Association for Computing Machinery}, \bibinfo{address}{New York, NY, USA}, Article \bibinfo{articleno}{473}, \bibinfo{numpages}{19}~pages.
\newblock
\showISBNx{9781450391573}
\urldef\tempurl%
\url{https://doi.org/10.1145/3491102.3501909}
\showDOI{\tempurl}


\bibitem[Wang et~al\mbox{.}(2020a)]%
        {wang2020human}
\bibfield{author}{\bibinfo{person}{Dakuo Wang}, \bibinfo{person}{Elizabeth Churchill}, \bibinfo{person}{Pattie Maes}, \bibinfo{person}{Xiangmin Fan}, \bibinfo{person}{Ben Shneiderman}, \bibinfo{person}{Yuanchun Shi}, {and} \bibinfo{person}{Qianying Wang}.} \bibinfo{year}{2020}\natexlab{a}.
\newblock \showarticletitle{From human-human collaboration to Human-AI collaboration: Designing AI systems that can work together with people}. In \bibinfo{booktitle}{\emph{Extended abstracts of the 2020 CHI conference on human factors in computing systems}}. \bibinfo{pages}{1--6}.
\newblock


\bibitem[Wang et~al\mbox{.}(2020b)]%
        {haicollab}
\bibfield{author}{\bibinfo{person}{Dakuo Wang}, \bibinfo{person}{Elizabeth Churchill}, \bibinfo{person}{Pattie Maes}, \bibinfo{person}{Xiangmin Fan}, \bibinfo{person}{Ben Shneiderman}, \bibinfo{person}{Yuanchun Shi}, {and} \bibinfo{person}{Qianying Wang}.} \bibinfo{year}{2020}\natexlab{b}.
\newblock \showarticletitle{From Human-Human Collaboration to Human-AI Collaboration: Designing AI Systems That Can Work Together with People}. In \bibinfo{booktitle}{\emph{Extended Abstracts of the 2020 CHI Conference on Human Factors in Computing Systems}} (Honolulu, HI, USA) \emph{(\bibinfo{series}{CHI EA '20})}. \bibinfo{publisher}{Association for Computing Machinery}, \bibinfo{address}{New York, NY, USA}, \bibinfo{pages}{1–6}.
\newblock
\showISBNx{9781450368193}
\urldef\tempurl%
\url{https://doi.org/10.1145/3334480.3381069}
\showDOI{\tempurl}


\bibitem[Wu et~al\mbox{.}(2022)]%
        {Wu_2022}
\bibfield{author}{\bibinfo{person}{Xingjiao Wu}, \bibinfo{person}{Luwei Xiao}, \bibinfo{person}{Yixuan Sun}, \bibinfo{person}{Junhang Zhang}, \bibinfo{person}{Tianlong Ma}, {and} \bibinfo{person}{Liang He}.} \bibinfo{year}{2022}\natexlab{}.
\newblock \showarticletitle{A survey of human-in-the-loop for machine learning}.
\newblock \bibinfo{journal}{\emph{Future Generation Computer Systems}}  \bibinfo{volume}{135} (\bibinfo{date}{oct} \bibinfo{year}{2022}), \bibinfo{pages}{364--381}.
\newblock
\urldef\tempurl%
\url{https://doi.org/10.1016/j.future.2022.05.014}
\showDOI{\tempurl}


\bibitem[Yampolskiy(2019)]%
        {unexplainable}
\bibfield{author}{\bibinfo{person}{Roman~V. Yampolskiy}.} \bibinfo{year}{2019}\natexlab{}.
\newblock \bibinfo{title}{Unexplainability and Incomprehensibility of Artificial Intelligence}.
\newblock
\newblock
\urldef\tempurl%
\url{https://doi.org/10.48550/ARXIV.1907.03869}
\showDOI{\tempurl}


\bibitem[Zhang et~al\mbox{.}(2019)]%
        {https://doi.org/10.48550/arxiv.1909.09906}
\bibfield{author}{\bibinfo{person}{Ruohan Zhang}, \bibinfo{person}{Faraz Torabi}, \bibinfo{person}{Lin Guan}, \bibinfo{person}{Dana~H. Ballard}, {and} \bibinfo{person}{Peter Stone}.} \bibinfo{year}{2019}\natexlab{}.
\newblock \bibinfo{title}{Leveraging Human Guidance for Deep Reinforcement Learning Tasks}.
\newblock
\newblock
\urldef\tempurl%
\url{https://doi.org/10.48550/ARXIV.1909.09906}
\showDOI{\tempurl}


\bibitem[Zoom(2011)]%
        {zoom}
\bibfield{author}{\bibinfo{person}{Zoom}.} \bibinfo{year}{2011}\natexlab{}.
\newblock \bibinfo{title}{One platform to connect}.
\newblock
\newblock
\urldef\tempurl%
\url{https://zoom.us}
\showURL{%
\tempurl}


\bibitem[Şahin et~al\mbox{.}(2007)]%
        {Planningaffordance}
\bibfield{author}{\bibinfo{person}{Erol Şahin}, \bibinfo{person}{Maya Çakmak}, \bibinfo{person}{Mehmet~R. Doğar}, \bibinfo{person}{Emre Uğur}, {and} \bibinfo{person}{Göktürk Üçoluk}.} \bibinfo{year}{2007}\natexlab{}.
\newblock \showarticletitle{To Afford or Not to Afford: A New Formalization of Affordances Toward Affordance-Based Robot Control}.
\newblock \bibinfo{journal}{\emph{Adaptive Behavior}} \bibinfo{volume}{15}, \bibinfo{number}{4} (\bibinfo{year}{2007}), \bibinfo{pages}{447--472}.
\newblock
\urldef\tempurl%
\url{https://doi.org/10.1177/1059712307084689}
\showDOI{\tempurl}
\showeprint{https://doi.org/10.1177/1059712307084689}


\end{thebibliography}

\appendix

\end{document}